\documentclass[10pt]{article}

\usepackage[utf8]{inputenc}
\usepackage[margin = 1in]{geometry}
\usepackage{hyperref}
\usepackage{amsmath}
\usepackage{pdfpages}
\usepackage{epstopdf}

\newcommand{\Rb}{\mathbb{R}}

\usepackage{quiver}
\usepackage{subcaption}
\usepackage{booktabs}

\begin{document}

\title{An Algebraic Framework for Structured Epidemic Modeling}

\author{Sophie Libkind, Andrew Baas, Micah Halter, Evan Patterson, and James Fairbanks
}

\maketitle{}

\begin{abstract}
Pandemic management requires that scientists rapidly formulate and analyze epidemiological models in order to forecast the spread of disease and the effects of mitigation strategies. Scientists must modify existing models and create novel ones in light of new biological data and policy changes such as social distancing and vaccination. Traditional scientific modeling workflows detach the structure of a model---its submodels and their interactions---from its implementation in software. Consequently, incorporating local changes to model components may require global edits to the code-base through a manual, time-intensive, and error-prone process. We propose a compositional modeling framework that uses high-level algebraic structures to capture domain-specific scientific knowledge and bridge the gap between how scientists think about models and the code that implements them. These algebraic structures, grounded in applied category theory, simplify and expedite modeling tasks such as model specification, stratification, analysis, and calibration. With their structure made explicit, models also become easier to communicate, criticize, and refine in light of stakeholder feedback.

\end{abstract}

\section{Introduction}

The basic principles of epidemics originate with the Kermack-McKendrick model, which established the paradigm of compartmental models to predict how an infectious disease spreads through a population over time. Yet, the COVID-19 pandemic has created a pressing need to expedite the development of new and customized epidemiological models. For instance, the Institute for Health Metrics and Evaluation (IHME), an independent global health research center at the University of Washington, has been collecting data, running simulations, and publishing findings throughout the COVID-19 pandemic. In April of 2020, just two months after the first confirmed US cases of COVID-19, the IHME was comparing different epidemiological models according to the accuracy of their forecasts~\cite{friedmanPredictivePerformanceInternational2021}. The models were summarized in natural language and compared mathematically via model outcomes derived by simulation. Such summaries do not permit a precise comparison of model \emph{structure}, which is essential to understanding how modeling assumptions affect model outcomes. Inadequate model representations limit the speed and precision with which organizations like the IHME can respond to emerging pandemics.

We present and exemplify an approach to developing scientific modeling software in which the high-level structure of a model is visible and manipulable in its software implementation. Mathematical models of scientific systems involve rich, structured knowledge that scientists leverage to communicate, iterate, and validate their models. However, this knowledge is obscured from computer systems when models are implemented using low-level computational primitives rather than by software tools that mirror the scientific concepts. Our approach has two phases: (1) formalizing the mathematics of high-level structures that recur in scientific models and (2) implementing these mathematical formalisms directly in software. Our approach contrasts with the traditional implementation of scientific models, because it treats the structure of scientific models as primary which in turn prioritizes the expertise of domain scientists and modelers during the development, adjustment, and analysis of models.  As an example of our approach, we formalize the specification of a compositional or stratified epidemiological model using the mathematics of applied category theory, and we demonstrate its implementation in software.\footnote{Code that demonstrates our software system and reproduces the examples in this manuscript is available on GitHub at \url{https://github.com/AlgebraicJulia/Structured-Epidemic-Modeling/}.}

Many aspects of model structure can be formalized as algebraic structure using mathematics from category theory. For example, we present a generalized addition and a generalized product for composing models. These operators are grounded in category theoretic concepts including copresheaves, structured cospans, and pullbacks. These concepts underlie our software design which automates the implementation of these operations. 
When specifying a composite model, a sharp distinction is drawn between the \emph{syntax of composition} dictating how the subsystems interact and the \emph{semantics of composition} assigning concrete mathematical models to the subsystems. This separation promotes generality and flexibility in modeling as the same syntax can have several different semantics. In this paper, we will see that Petri nets with mass action kinetics and, more generally, ordinary differential and delay differential equations, can all serve as semantics for the same syntax. Furthermore, translations between these and other types of models can be formalized using concepts from category theory. Unlike the textual syntax of conventional programming languages and data formats, our syntax is algebraic and diagrammatic in nature. This language makes the informal diagrams used by scientists and engineers mathematically rigorous and directly computable. Furthermore, the syntax itself becomes a combinatorial data structure that can be algorithmically manipulated. \textit{Undirected wiring diagrams}, which form an algebraic structure called an \emph{operad}, are the main syntax treated in this paper, although other syntaxes, such as directed wiring diagrams, can be similarly applied \cite{vagner2015}, \cite{libkind2021operadic}.

Although all scientists intuitively understand that complex models are assembled from simpler ones, such decompositions are rarely made explicit in modeling practice. Our operadic approach to compositional modeling makes the modular structure transparent and rigorous. As a result, experts in distinct fields can develop submodels in parallel, particular submodels can easily be replaced without affecting others, and programming errors are reduced because the software handles the bookkeeping needed to assemble the composite model. All of this serves to accelerate the modeling process, which is critical in emerging scenarios where existing models must be rapidly adapted to new scientific contexts.

Because programming simulators is labor intensive and error-prone, many existing software systems present higher-level interfaces for modelers to formulate and analyze models. They are often specific to a particular scientific domain or theoretical class of models. For example, Stan is used for analyzing data with probabilistic models \cite{stan}, Kappa for rule-based modeling of biochemical systems \cite{kappa}, Copasi for simulating mechanistic models in systems biology \cite{copasi}, \texttt{pomp} for partially observed Markov processes \cite{king2016pomp}, NetLogo for agent-based modeling \cite{netlogo}, and Berkeley Madonna for the graphical construction of complex mathematical models. When several scientific tools evolve in the same domain, they may undergo a process of standardization leading to new interchange formats, such as the Systems Biology Markup Language~\cite{sbml}, that can be consumed by many tools. While such tools and formats might enable rapid model formulation and analysis at a certain scale, they often lack features like hierarchical composition needed to reliably specify large-scale models. For example, in the absence of other organizing abstractions, Petri net models rapidly devolve into an incomprehensible web of criss-crossing transitions. Existing software is also typically bespoke, designed for a specific field and not around common abstractions that recur throughout mathematics. Building software around such abstractions facilitates the construction of models that cross disciplinary boundaries while allowing for greater generality and code reuse.

Our software system encompasses a range of tasks in the scientific modeling workflow including model formulation, simulation, analysis, and comparison. The paper is organized along these lines. Section \ref{sec:compositional} presents the compositional approach to model specification, explaining how undirected wiring diagrams are a syntax for composing open Petri nets as well as more general differential equation models. Subsection~\ref{sec:advantages} highlights the computational and theoretical advantages of this approach.  Section~\ref{sec:stratification} demonstrates how maps between Petri nets can encode domain-specific type systems, which are useful for building stratified models and verifying that models do not violate established theory. Finally, Section \ref{sec:analysis} discusses calibration and sensitivity analysis via interoperation with other tools, emphasizing visualization and the tight feedback loop enabled by high-level model specifications. 

\section{Compositional methods of model specification}\label{sec:compositional}

\subsection{Structured multicospans of Petri nets for compartmental models}\label{sec:petrinets}

In this section, we exemplify the categorical approach to compositional modeling by giving the three components of such a framework: (1) a model semantics, for specifying concrete mathematical models of systems, (2) a composition syntax, for specifying interactions between systems, and (3) a composition rule, for specifying how to compose chosen concrete models according to a given syntactic term. We emphasize that the composition syntax is a logical syntax, in that the composition syntax axiomatizes the rules of well-formed expressions that define a composition of models. In contrast, the model semantics is a denotational semantics which assigns meaning to the models in a composition. We do not mean semantics in the sense of observed, real-world phenomena and such an interpretation is outside of the scope of this work.

\begin{figure}
    \centering
    \includegraphics[width = 0.9\textwidth]{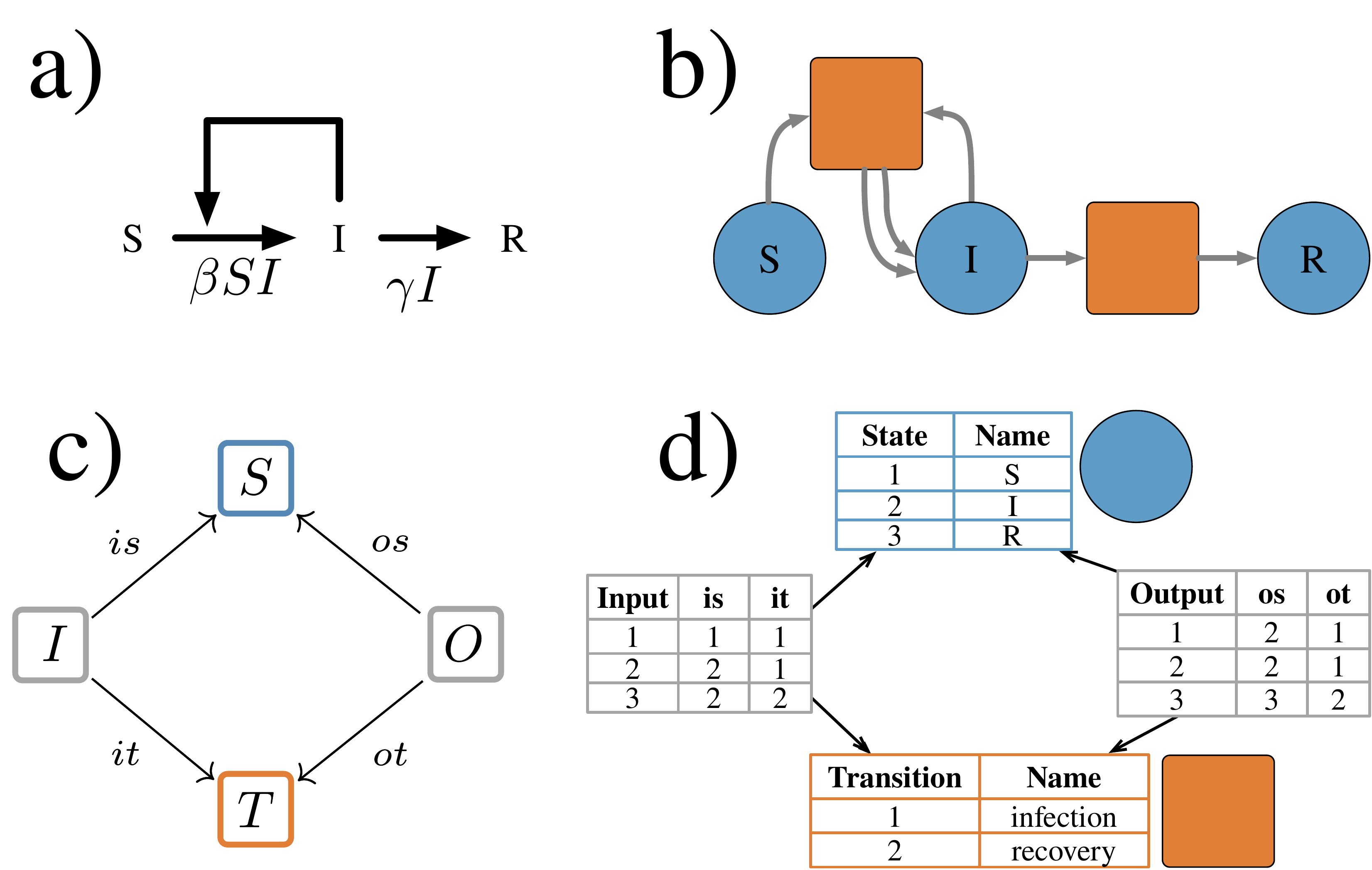}
    \caption{(a) An informal visualization of a compartmental SIR model. The horizontal arrows represent processes in which susceptible people become infected and infected people recover. The single feedback arrow indicates that the rate of infection  depends on the size of the infected population.  (b) The formalization of the comparmental model in (a) as a whole-grain Petri net. It has  three \emph{species} (depicted as labeled blue circles) representing a susceptible, infected, and recovered population and two \emph{transitions} (depicted as orange squares) representing an infection process and a recovery process. The transition representing infection has two \emph{input arcs} and two \emph{output arcs}. These indicate that infection takes one susceptible and one infected individual and returns two infected individuals. The transition representing recovery has one input arc and one output arc. These indicate that recovery takes one infected individual and returns one recovered individual. (c) The diagram of finite sets and functions which define a whole-grain Petri net. $S$ is a finite set of \emph{species}, $T$ is a finite set of \emph{transitions}, $I$ is a finite set of \emph{input arcs}, and $O$ is a finite set of \emph{output arcs}. (d) The database view of the whole-grain Petri net depicted in (b). The database view gives a table for each finite set in the diagram. Table rows represent the elements of the set and when applicable table columns represent the functions out of the set. }
    \label{fig:Petri}
\end{figure}

For the model semantics, we use Petri nets in a form called \emph{whole-grain Petri nets} \cite{kock2020}, defined by the diagram of finite sets and functions shown in Figure~\ref{fig:Petri}(c). Thus, a whole-grain Petri net consists of a finite set of \emph{places} or \emph{species} $S$, a finite set of \emph{transitions} $T$, and spans $S \xleftarrow{is} I \xrightarrow{it} T$ and $S \xleftarrow{os} O \xrightarrow{ot} T$ defining the \emph{input} and \emph{output} arcs between states and transitions. A span is similar to a ``multirelation" in that pairs of elements may be related with multiplicity. Multiplicity is important for transitions that input or output multiple tokens of the same species, such as an infection transition that yields two infected individuals as output. As an example, consider the SIR model in Figure~\ref{fig:Petri}(b). This Petri net has three places $S = \{S, I, R\}$ corresponding to susceptible, infected, and recovered populations and two transitions corresponding to infection and recovery. The infection transition is the target of two input arcs --- one whose source is the place $S$ and one whose source is the place $I$ --- and the source of two output arcs --- both whose target is the place $I$. The recovery transition is the target of a single input arc and the source of a single output arc.

Petri nets are closed systems, meaning that they are isolated from interaction with other systems. Non-compositional modeling approaches focus on explicitly defining and implementing closed systems. However, Petri nets arising in practice can comprise hundreds \cite{coexist} or thousands \cite{wu2021} of states and transitions, making them unwieldy both to conceptualize and to implement. In contrast, our compositional modeling approach capitalizes on the tendency of real-world systems to coexist in richly-structured ecosystems and enables the development and assembly of open-system models.

\begin{figure}
    \centering
    \includegraphics[width=\textwidth]{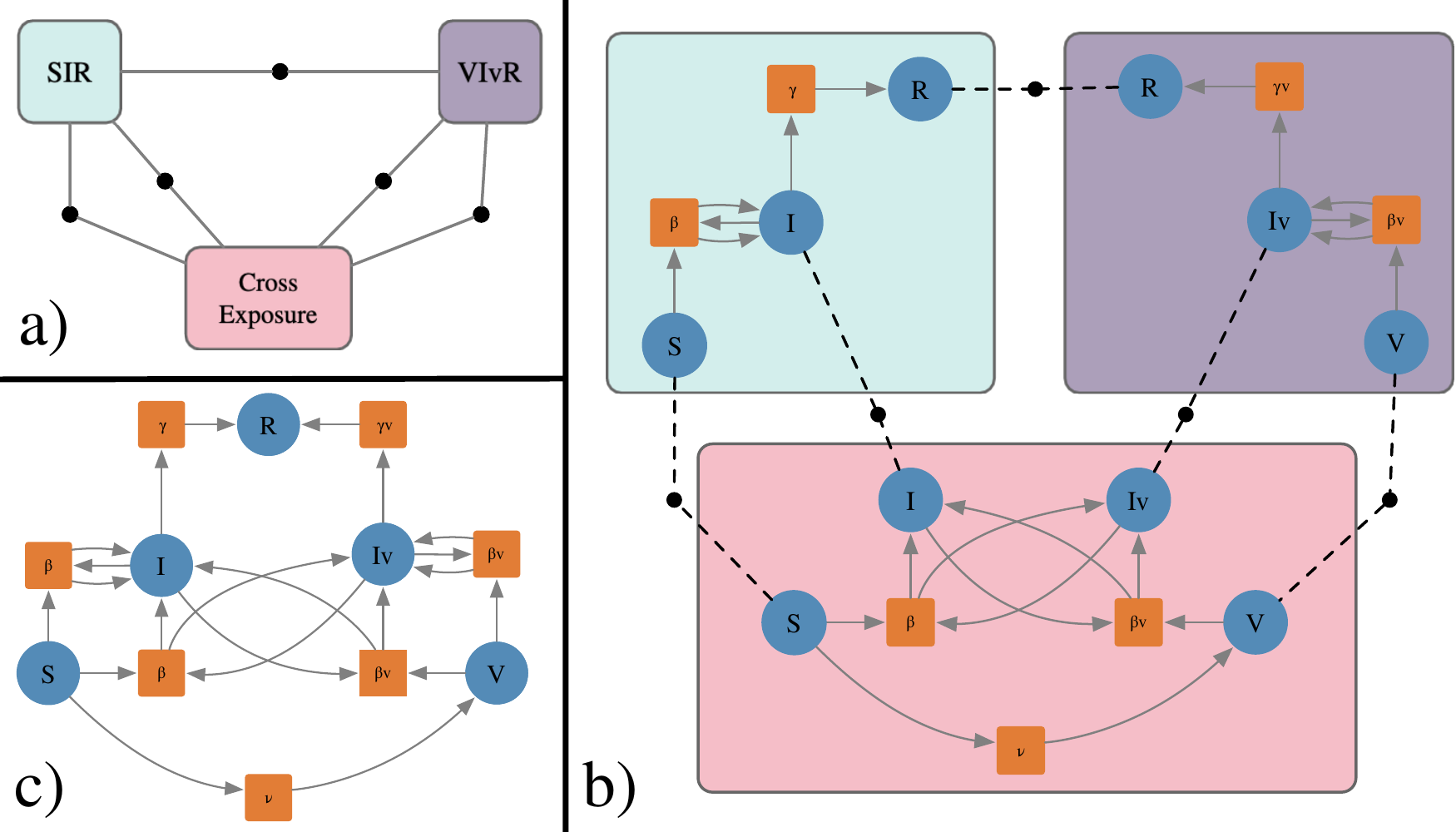}
    \caption{An example of a Petri net for a compartmental model specified as a composite of submodels. (a) The syntax for composition is given by an undirected wiring diagram which represents the high level design of the subsystems and their interactions. (b) The component submodels are given as open Petri nets. Starting with the submodel in blue and moving clockwise they are a submodel isomorphic to the classic susceptible-infected-recovered (SIR) disease spread model for unvaccinated people; a submodel with presumably lower transmission rates isomorphic to the SIR disease spread model for vaccinated people; and  a submodel specifying the interactions between the two populations, namely the vaccination process itself and transmission between vaccinated and unvaccinated people. While in this example the variable names for the places are aligned between the submodels, this need not be the case in generic compositions. The wires in the UWD (represented in (b) by the dashed lines) explicitly allow for component submodels to apply different naming conventions and preserve the independence of submodels. (c) The AlgebraicJulia software computes the  composite model from the high-level design and the specification of components.}
    \label{fig:sviivr}
\end{figure}

Structured cospans \cite{baez2020structured} and decorated cospans \cite{fong2015decorated} are formalisms for turning closed model semantics into open model semantics, in which systems can interact along specified interfaces. Although structured cospans are applicable to a variety of systems, mathematically and in our implementation, we restrict our discussion to the important case of Petri nets. A \emph{structured multicospan of Petri nets}, or \emph{open Petri net} (cf. \cite{baezOpenPetriNets2020}), is a whole-grain Petri net~\cite{kock2020} together with a list of finite sets $A_1, \dots, A_n$ and functions $A_1 \to S, \dots, A_n \to S$. The sets $A_i$, called the \emph{feet} of the structured multicospan, define an interface for the open Petri net. The functions $A_i \to S$, called the \emph{legs} of the structured multicospan, select the places of the  Petri net which are exposed through the interface. The more standard notion of \emph{structured cospan} is the special case of $n=2$ legs. The extra flexibility afforded by multicospans is useful in practice. In particular, open Petri nets with arbitrary numbers of legs can be composed using the graphical syntax of undirected wiring diagrams (UWDs) \cite{spivak2013}, which are a generic graphical syntax for composing relations, database tables, structured multicospans, and other undirected systems. An UWD consists of a set of \textit{boxes}, a set of \textit{ports}, and a set of \textit{junctions}. Each port is assigned to a box and wired to a junction. Figure~\ref{fig:sviivr}(a) depicts a UWD with three boxes, ten ports, and five junctions. A port assigned to the SIR box and a port assigned to the VIvR box are both wired to the top-most junction. Likewise, two junctions connect two ports assigned to the SIR box with two ports assigned to the cross exposure box, and two junctions connect two ports assigned to the VIvR box with two ports assigned to the cross exposure box.

In Figure~\ref{fig:sviivr}, we present a compartmental model for viral dynamics that accounts for vaccination as the composition of three concrete submodels corresponding to (1) a disease spread model for unvaccinated people, (2) a disease spread model for vaccinated people, and (3) a cross exposure model of the interactions between the two populations.
The open Petri nets for these three primitive subsystems are shown in Figure~\ref{fig:sviivr} (b). The composition syntax is given by the UWD in Figure \ref{fig:sviivr}(a), which has three boxes, ten ports, and five junctions.
The systems compose by identifying places that are connected in the UWD. For example, the recovered populations, labeled $R$, in the SIR and VIvR models are identified in the composite model. The resulting composite model is shown in Figure~\ref{fig:sviivr} (c). 

In mathematical terms, UWDs form an \emph{operad} constituting the composition syntax, and the structured multicospans of Petri nets (strictly speaking, their isomorphism classes) form an \emph{operad algebra} of the UWD syntax. This operadic framework has many advantages.
In addition to the modular model specification strategy exemplified in Figure~\ref{fig:sviivr}, the algebra  enables a hierarchical model specification strategy in which a submodel may itself be the composite of still more primitive submodels.  A syntax for hierarchical modeling is given later in Figure~\ref{fig:malaria}(f).
The operadic framework also enables a mathematically rigorous divide-and-conquer workflow by designating subsystems that can be developed and refined in parallel. 
Updates to submodels do not affect others except through explicitly represented changes to the composition syntax.
Furthermore, the syntax provides an opportunity to build assumptions and domain-specific knowledge directly into models and can be used to identify properties or appropriate sampling algorithms of the composite model. These advantages are discussed further in Section~\ref{sec:advantages}.

The Julia packages AlgebraicPetri and AlgebraicDynamics directly implement the operadic approach described in this section and its extensions in Section~\ref{sec:malaria}. These packages enable modelers to create executable code for composite models that reflect the modular and hierarchical structure of real-world systems~\cite{libkind2021operadic}.

\subsection{Mass action kinetics for open Petri nets}\label{sec:mass_action}

\newcommand{\Nat}{\mathbb{N}}

A Petri net is a combinatorial description of a dynamical process. The graphical representation and network topology of Petri nets can be analyzed to infer structural properties of the system. However, behavioral analyses often require explicit model simulations. These simulations can be computed using discrete sampling algorithms, such as Gillespie's direct method or tau-leaping, or by interpreting a Petri net as an ordinary differential equation (ODE) and applying standard numerical integration techniques. In this section we focus on the latter method, which allows for integration with the calibration and analysis toolkits described in Section~\ref{sec:analysis}.

Following Baez and Pollard \cite[Definition 13]{baezCompositionalFrameworkReaction2017}, we associate an ODE to a Petri net by applying the law of mass action, which states that transitions consume inputs and produce outputs at rates proportional to the product of their input concentrations.  To illustrate, define the function $p:S \to \Nat^T$ so that $p(s)$ is the multiset of transitions producing the species $s$. Likewise, define  $r: T\to \Nat^S$ so that $r(t)$ is the multiset of species that are inputs to the transition $t$. Its weighted preimage $r^{-1}: S\to \Nat^T$ maps a species $s$ to the multiset of transitions for which it is an input. Each species $s$ in the Petri net is assigned a variable $u_s$ in the ODE. The transitions define the following vector field on the state space $\Rb^S$: 
\begin{equation} \label{eqn:mak}
    \dot u_s = \sum_{t \in p(s)} \phi_t - \sum_{t\in r^{-1}(s)} \phi_t,
    \qquad\text{where}\qquad
    \phi_t := \beta_t \prod_{s\in r(t)} u_s
\end{equation}
and $\beta_t$ is the rate constant associated with transition $t$. These equations define the standard interpretation of Petri nets as systems of ODEs governing chemical reaction networks. Note that the multiset multiplicities represent the stoichiometric coefficients in the chemical reaction interpretation of the Petri net. 

When applicable, defining ODEs by Petri nets or composites of Petri nets has significant software advantages. Meaningful, local changes to a Petri net, such as substituting submodels in a composition or adding a single species or transition, often lead to nontrivial, global changes to the corresponding ODE that affect many variables and terms. For example, analysis tools that observe, report, and calculate properties of state variables throughout a simulation often refer to a single variable in many places. Therefore, adding or removing a state variable to the system requires making many coordinated changes to the code. This design pattern results in software where local changes to the mathematical model require global changes to the software implementation. In contrast, local changes to a Petri net model requires only local changes to its software implementation in AlgebraicPetri. The transformation of a Petri net into an ODE via the law of mass action automatically and accurately translates these local changes to the Petri net model into global changes to the corresponding ODE. This automation can accelerate the modeling cycle---such as for making modifications in response to new information or testing for policy robustness---which is critical when responding to urgent situations.

\subsection{Composition of general differential equation models}\label{sec:malaria}

\begin{figure}
    \centering
    \includegraphics[width=0.99\textwidth]{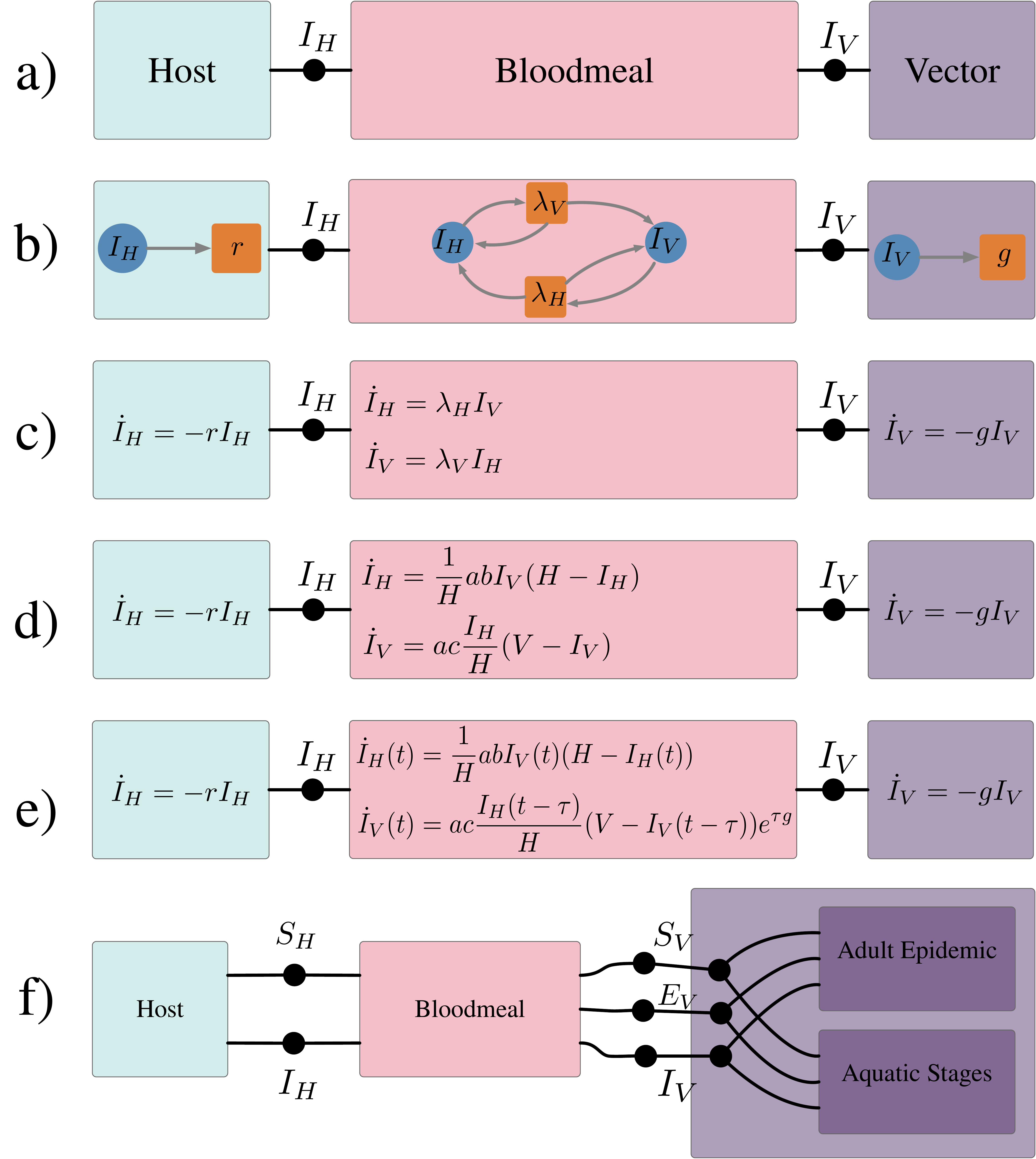}
    \caption{The UWDs  in (a) and (f) define syntaxes for composing submodels so that the result is  a model for the spread of a vector-borne pathogen. In (a) the junctions represent a shared population of infected hosts ($I_H$) and a shared population of infected vectors ($I_V$). In (f) there are additional junctions for the shared populations of susceptible hosts ($S_H$), susceptible vectors ($S_V$), and infected but not yet infectious vectors ($E_V$). The syntax in (f) corresponds to several of the modules presented in~\cite{wu2021}. For the syntax in (a), different choices of concrete submodels produce different composite models of varying types and complexity. Several such choices are depicted in (b-e). (b) presents the components as Petri nets and (c) gives ODEs which are the law of mass action applied to the components in (b).  (d) represents a Ross-Macdonald ODE model while (e) incorporates a delay which accounts for the incubation period for the pathogen in vectors.  In (d) and (e) the parameter $a$ is the biting rate, $b$ is the infection efficacy for hosts, $c$ is the infection efficacy for vectors, $H$ is the total host population and $V$ is the total vector population. }
    \label{fig:malaria}
\end{figure}

Mass action kinetics are often insufficient to simulate complex dynamical processes like those found in biology, ecology, and epidemiology. In this section, we show how the composition method for Petri nets generalizes to composition methods for models of different types, in particular to models explicitly defined by ordinary differential equations (ODEs) or delay differential equations (DDEs).

The composition of Petri nets described in Section~\ref{sec:petrinets} is an example of the categorical formalism of operads and operad algebras, which equip visual grammars with the rigor of  algebraic equations. The theme of the operadic approach is to explicitly and independently describe the syntax of composition---how subsystems interact---and the semantics of composition---particular choices of component models for each subsystem. The example of composing Petri nets, in which the syntax is given by an undirected wiring diagram and the semantics by open Petri nets, is one of many examples of the operadic approach to modeling.

Modeling vector-borne pathogens, such as malaria, is an exemplar context for the operadic approach because the pathogen dynamics naturally decompose into distinct scientific domains such as epidemiology and entomology. The Ross-Macdonald class of equational models highlights the decomposition of a vector-borne pathogen epidemic into three subsystems: pathogen dynamics in the vector (mosquito) population, pathogen dynamics in the host (human) population, and the dynamics of pathogen transference in the bloodmeal~\cite{smith2014}. Syntactically, this composition is defined by a UWD with three boxes corresponding to the three subsystems and with junctions corresponding to populations involved in multiple processes. Figure~\ref{fig:malaria}(a) defines this UWD. To this composition pattern, we apply submodels of increasing complexity and of different types, namely Petri nets in Figure~\ref{fig:malaria}(b), ODEs in Figure~\ref{fig:malaria}(c,d), and DDEs in Figure~\ref{fig:malaria}(e). The composition for Petri nets was defined in Section~\ref{sec:petrinets}. The ODE and DDE submodels compose by identifying variables connected in the UWD and summing the rates of change for identified variables. For example, the result of composing the three ODE submodels in Figure~\ref{fig:malaria}(d) according to the given composition pattern is $$\dot I_H = ab \frac{I_V}{H}(H - I_H) - rI_H, \quad  \dot I_V = ac \frac{I_H}{H}(V - I_V) - gI_V.$$

Just as the law of mass action defines a transformation from Petri nets to  ODEs, there is also a transformation from ODEs to DDEs giving a trivial dependence on the history. These transformations allow modelers to use different model types to specify the submodels of a composite. For example, in Figure~\ref{fig:malaria}(e) the submodels for the pathogen dynamics in hosts and vectors are given by ODEs and in contrast the bloodmeal model is given by a DDE. The composite model is derived by translating the ODE submodels into DDE models and then composing the DDE models with the result being akin to the Sharpe-Lotka model \cite{sharpe1978}. This formal process gives domain experts the freedom to choose model classes that best fit their field.

\subsection{Discussion of compositional model specification}\label{sec:advantages}
In this section, we have presented a compositional approach to modeling that is grounded in the mathematics of applied category theory and implemented directly in software. We conclude with a discussion of this framework.

{\bf Advantages to the engineering process.} Engineering is a process that involves taking a theoretical description of a model and developing software that can simulate, calibrate, and analyze the model. A model description is often  informally compositional and in traditional software this structure is implicit in the code. In contrast, software packages based on the operadic approach to modeling make this structure explicit and disambiguate the process of turning the mathematical specification of a model into the code that implements it. As a result, engineers have a mathematically grounded divide-and-conquer approach to select, implement, and iteratively develop submodels. This process is also hierarchical as the categorical formalism implies that a submodel may itself be the composition of still more primitive models. Furthermore, as discussed in Section~\ref{sec:mass_action}, the implementation of the categorical framework can reduce code complexity and errors, since local changes to models correspond to local changes in the code base. 
    
{\bf Advantages to the scientific process.}
    The scientific process relies on transparent communication and critique of models, and a common problem is that the shortest description of a model is the code itself rather than the theoretical model description. While the code is precise, it is often not easily or efficiently understood, even by proficient programmers. 
    Strategies such as the Overview, Design Concepts, and Details (ODD) Protocol alleviate these strains on the scientific process by establishing documentation conventions and encouraging the assumptions and theoretical underpinnings of a model to be made rigorous and communicable~\cite{grimm2010odd}.
    Our framework takes this strategy a step further by grounding the model description in an algebraic structure. The visual diagrams used to communicate models, such as those in Figures~\ref{fig:sviivr} and~\ref{fig:malaria}, then become rigorous enough to be unambiguously translated into code.
    By making a model's theoretical formulation more visible and more tightly bound to its software implementation, the compositional approach helps modelers identify components or interactions that are unnecessarily complicated, do not properly reflect domain knowledge, or depend upon unreasonably strong assumptions.

    The compositional framework for modeling also streamlines the scientific process by prioritizing the independence of submodels. Submodels can be efficiently tested and substituted without affecting other submodels in a composite.  For example,  Figure~\ref{fig:malaria}(c-e) exemplifies updating the bloodmeal submodel without affecting the host and vector submodels. This feature is practical for (1) model formulation, in which parsimonious but empirically adequate models are found by testing different combinations and complexities of submodels, and (2) policy making, where it is important that policies be robust to variations in submodels and other modeling assumptions. Works such as \cite{citron2021} and \cite{wu2021} demonstrate the importance of testing multiple combinations of submodels. The categorical approach and its implementation in software disambiguates and assists this process.
    
{\bf Theoretical advantages.} The compositional framework also provides theoretical advantages to scientific modeling. 
    Because the syntax and semantics of composition are explicitly and independently represented, the composition syntax is a venue for exchanging expert knowledge, while the choice of submodels can be left to specific domain experts or a model selection process. For example, the syntax proposed in Figure~\ref{fig:malaria}(f) asserts that the submodels for the vector dynamics must include a susceptible, an infected but not yet infectious, and an infectious population. It also specifies that the model of vector dynamics is broken down into submodels for the aquatic stages and for the epidemic in adults. Additionally, the composition syntax can be analyzed for mechanistic or causal dependencies. For instance, the syntax given in Figure~\ref{fig:malaria}(f) expresses that the host and vector population can only affect each other through the bloodmeal. 
    
    Finally, mathematical properties of these compositional modeling frameworks translate directly into important consistency properties for model construction. The associativity and symmetry properties of operads and their algebras imply that the order of composing submodels does not affect the final model. Similarly, the functoriality of reinterpretation rules, such as the law of mass action, implies that reinterpretation and composition can be done in any order, which again does not affect the final result.


\newcommand{\epi}{\mathsf{infectious}}
\newcommand{\born}{{\mathsf{vector}\text{-}\mathsf{borne}}}

\section{Type systems for open Petri nets}\label{sec:stratification}
\subsection{Typed Petri nets}
Category theory emphasizes the importance of \textit{morphisms} or maps between mathematical objects. In this section, we demonstrate how morphisms between Petri nets can be used to define typed Petri nets.

\begin{figure}
    \centering
    \includegraphics[width=0.8\textwidth]{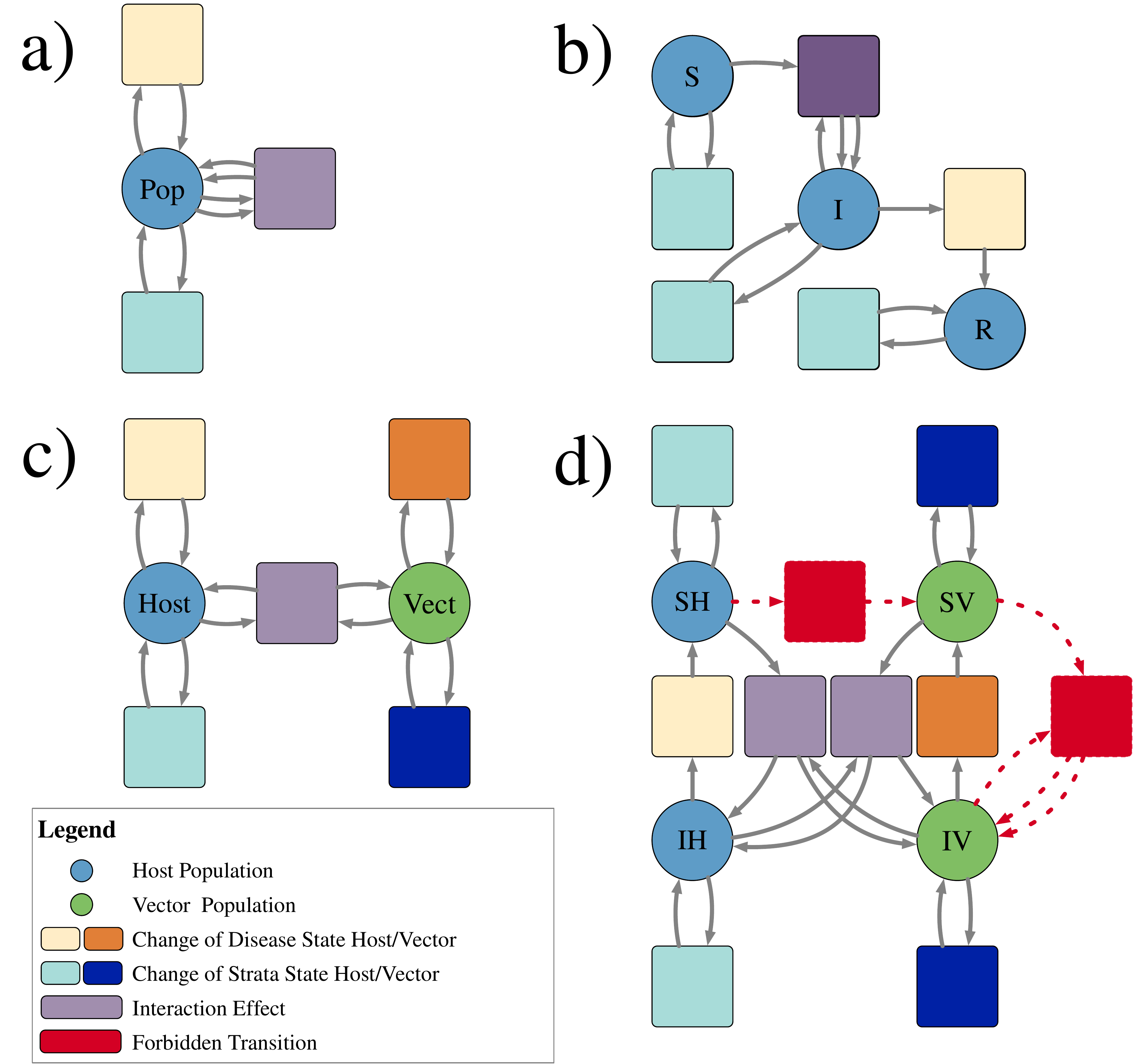}
    \caption{(a) A type system, $P_{\epi}$, representing Petri nets for infectious diseases. It has a single place type and three transition types corresponding to changes in infection status (yellow), changes between strata (blue), and interactions (purple). (b) An SIR disease model typed by $P_{\epi}$. Colors indicate the type of each place and each transition. For example, the transition mediating the infected and recovered population is of type ``change in disease status" while the three transitions from each species to itself is of type ``change in strata," since a change in strata does not affect an individual's infection status. (c) A type system, $P_{\born}$, representing Petri nets for vector-borne diseases. It has two place types corresponding to the vector and host populations and specifies that interactions can only occur between vectors and hosts. (d) An SIS disease model typed by $P_{\born}$. There are two places typed by the host population representing a susceptible host population and an infected host population. Likewise for the vector population. The transitions indicated in red have no valid typing and thus are forbidden by the type system. The transition $S_H \to S_V$ would change the species of an individual, while the transition $I_V + S_V \to 2I_V$ would allow vectors to infect each other. Both of these transitions violate established scientific knowledge, and the type-system provides guardrails to prevent such modeling errors.}
    \label{fig:type_systems}
\end{figure}

Petri nets can represent domain-specific type systems. For example, the Petri net $P_\epi$ in Figure~\ref{fig:type_systems}(a) defines a type system for an infectious disease model. It consists of a single species type and three transition types corresponding to (1) spontaneous changes in infection status; (2) spontaneous changes between non-infection-related strata, such as movement between patches or changes in quarantine status; and (3) interactions between a pair of individuals. In contrast, the Petri net $P_\born$ depicted in Figure~\ref{fig:type_systems}(c) represents a type system for a vector-borne disease model and has two species types corresponding to the vector and host populations.

\begin{figure}
    \centering
    \includegraphics[width = 0.8\textwidth]{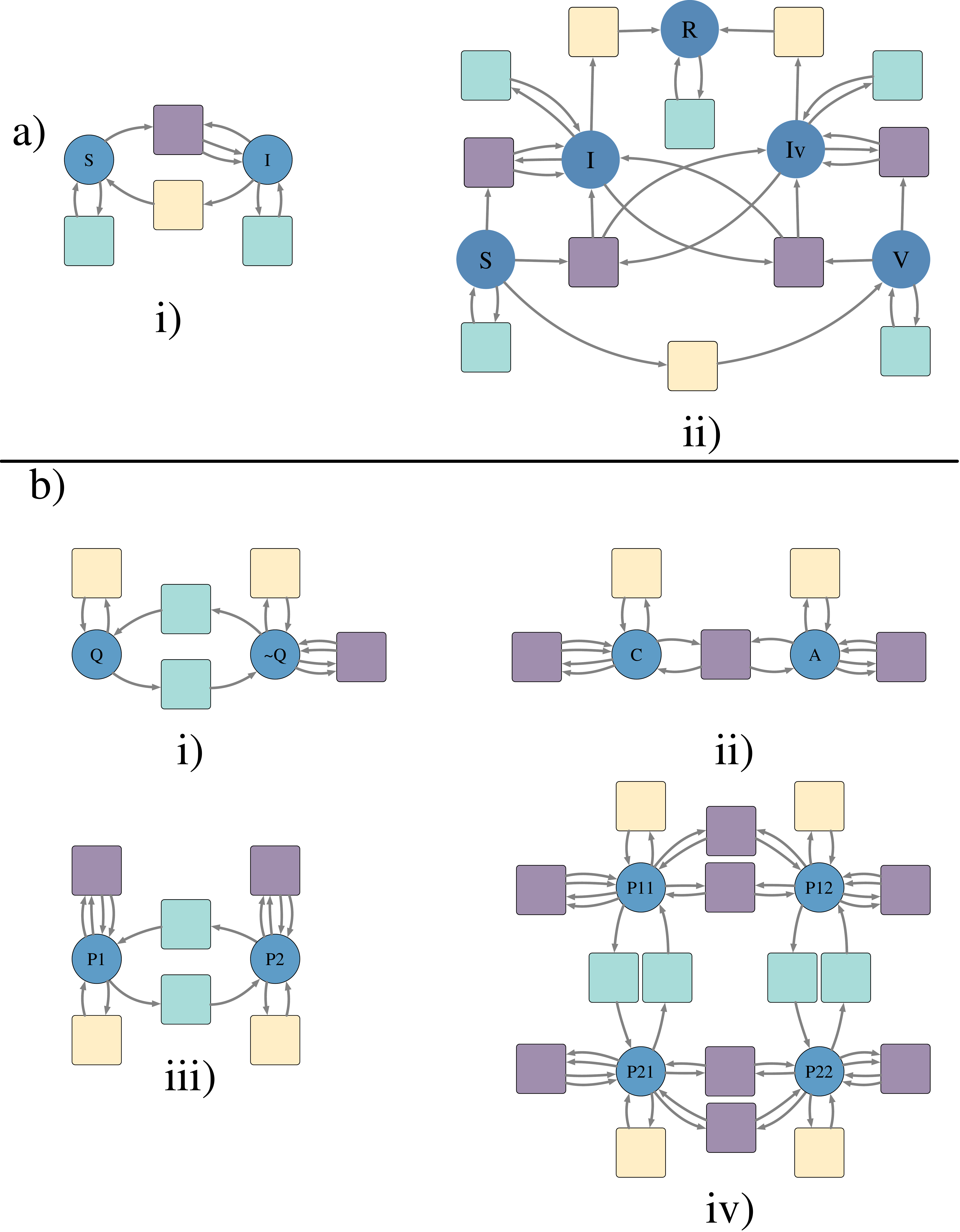}
    \caption{(a) A palette of disease models typed by $P_{\epi}$ including (a.i) a SIS model and (a.ii) a SVIIvR model which incorporates vaccination. (b) A palette of stratification schemes typed by $P_{\epi}$.  (b.i) Stratification by quarantine/isolation status. Interactions cannot  occur between individuals in the place representing the quarantining/isolating population. (b.ii) Stratification by age. Interactions can occur between children and adults. Although not pictured here, the map on arcs specifies if adults infect children or vice versa. Note that there are no transitions with type ``change in strata," representing that in this model children do not spontaneously become adults and vice versa.  (b.iii) The flux model for spatial dynamics. Each place represents a different patch. Individuals can move between patches, and interactions only occur between people in the same patch. (b.iv) The simple trip model of spatial dynamics. $P_{ij}$ represents a population that  is currently in patch $i$ and whose residence is patch $j$. Individuals can travel between patches but not change their residence. Interactions only occur between people currently in the same patch.  In both (a) and (b), transitions are colored to indicate the typing by $P_{\epi}$, the typed system depicted in Figure~\ref{fig:type_systems}(a).}
    \label{fig:palette}
\end{figure}

A morphism between Petri nets is a map of places, transitions, input arcs, and output arcs that preserves the arities of the arcs and respects the sources and targets of the arcs.\footnote{Such morphisms of Petri nets, also called \textit{etale maps}, are defined in ~\cite[Section 2.2]{kock2020}.} For example, the source of an input arc $i$ in the domain Petri net must map to the source of the arc to which $i$ is mapped. Given a Petri net $P_\mathsf{type}$ defining the type system, a \textit{typed Petri net} is a Petri net $P$ together with a morphism $\phi: P \to P_\mathsf{type}$.  Figure~\ref{fig:type_systems}(b) and Figure~\ref{fig:palette} give examples of Petri nets typed by $P_\epi$.

All of the Petri nets typed by a given type system form a \textit{slice category} of the category of whole-grain Petri nets. The mathematical features of slice categories guarantee important modeling features. First, typed Petri nets are practical for model checking. A Petri net typed by $P_\epi$  assigns each transition to be a spontaneous  change in infection, a spontaneous change in strata, or an interaction. The type of a transition must be consistent with the number of input and output arcs connected to it. For example, a typing by $P_\epi$ ensures that a transition with interaction type has two inputs and two outputs. Second, typed Petri nets facilitate high-level critiques of a model. For example, a model typed by $P_{\born}$ cannot incorporate vertical transmission from parents to offspring or sexual transmission in either hosts or vectors. This property may contradict known transmission pathways for a specific disease and thus motivate a revision of the model and the type system. Third, features of the type system may also directly translate into features of the typed Petri net. For example, because each transition in $P_\epi$ has the same number of inputs and outputs, any Petri net typed by $P_\epi$ conserves the total population over time. 

Typed Petri nets also provide guardrails for composing models.  A Petri net typed by $P_\born$ must assign each species to be of either  vector-type or of host-type thereby separating the vector and host populations. The type system also ensures that interactions only occur between vectors and hosts and that no vectors spontaneously become hosts and vice versa. For example in Figure~\ref{fig:type_systems}(d), the transition $S_H \to S_V$ which turns susceptible hosts into susceptible vectors is forbidden by the type-system. When composing open Petri nets typed by the same type system via the process described in Section~\ref{sec:petrinets}, we can add a constraint that identified species must have the same type. With this check in place, a host subpopulation will not be identified with a vector subpopulation during model composition. Furthermore, under this constraint, the composition of typed Petri nets retains a typing. 

Ultimately, a domain-specific typing can guarantee meaningful properties of a model and prevent novice users or automated systems from generating models that contradict common sense or domain expertise.

\subsection{Stratified compartmental models}

\begin{figure}
    \centering
    \includegraphics[width = 0.9\textwidth]{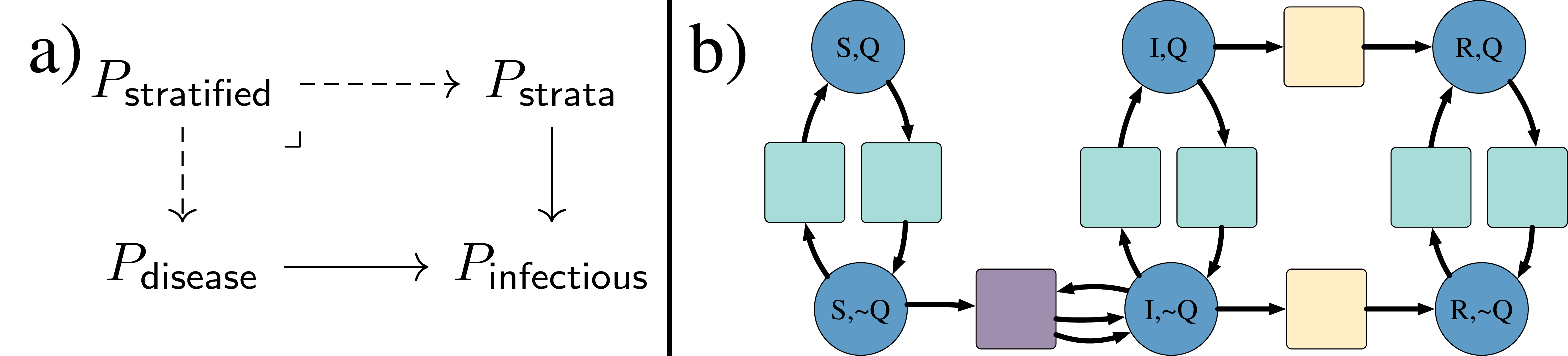}
    \caption{(a) A stratified model over $P_{\epi}$ is the pullback of a typed epidemiological model $P_{\mathsf{disease}} \to P_{\epi}$ and a typed stratification scheme $P_{\mathsf{strata}} \to P_{\epi}$. (b) The stratified model of the SIR model depicted in Figure~\ref{fig:type_systems}(b) and the quarantining model depicted in Figure~\ref{fig:palette}(b.i). }
    \label{fig:stratification}
\end{figure}

A recurring theme in scientific modeling is the importance of stratified models, in which local dynamics are reproduced in multiple strata and strata interact according to a specified scheme. For example, Citron et al~\cite{citron2021} compare stratified models defined by a choice of local epidemiological dynamics (SIR, SIS, or Ross-Macdonald) and a choice of stratification by location (the flux or simple trip models of metapopulation dynamics). Typed Petri nets offer a general methodology for stratifying models, which contrasts with the by-hand approach taken in~\cite{citron2021}. 

Consider two typed Petri nets, the unstratified \emph{disease model} $\phi: P \to P_\mathsf{type}$ and a \emph{stratification scheme} $\phi': P' \to P_\mathsf{type}$. The \textit{stratification} of $P$ by $P'$ is defined to be the Petri net with places (resp. transitions, input arcs, and output arcs) consisting of pairs of places (resp. transitions, input arcs, and output arcs) in $P$ and $P'$ which have the same type.  Because the morphisms $\phi$ and $\phi'$ respect the source and target maps from arcs to species and transitions, the source and target maps in the stratified model are well-defined. Figure~\ref{fig:stratification}(b) gives an example of stratifying the SIR model by a model of quarantine/isolation status. In the stratified model, the place $(S, Q)$ represents the susceptible and quarantining population while $(S, \sim Q)$ represents the susceptible and not quarantining population. The transition mediating the places $(S, \sim Q)$ and $(I, \sim Q)$ represents the infection transition in the SIR model and the interaction between non-quarantining people in the stratification scheme. Because these transitions are both of interaction type and mediate species of the same type,  they are paired in the stratified model. 
Figure~\ref{fig:palette} gives a palette of additional epidemiological models (SIS and SVIIvR) and  stratification schemes (quarantine status, age, the flux model of spatial dynamics, and the simple trip model of spatial dynamics). 

The categorical abstraction standardizes the definition of model stratification, and its implementation in AlgebraicPetri automates the construction of stratified models under the constraints of the expert-chosen type system. Because the size of stratified models grows quadratically with respect to the sizes of the component models, this framework streamlines the accurate implementation of stratified models as well as the clear communication and critique of stratified models via their components. Many of the advantages described in Section~\ref{sec:advantages} also apply to the categorical representation of model stratification.

Comparison with~\cite{citron2021} highlights the clarity and efficiency that our approach brings to the modeling workflow. The authors of~\cite{citron2021} investigate the choice of applying candidate models of movement  between subpopulations to disease models when calibrated to real-world data sets. They combine three standard disease models (SIR, SIS, and Ross-Macdonald) with two choices of movement dynamics (flux and simple trip). In the flux model, people relocate to different patches at fixed rates. In the simple trip model, people are assigned a home patch and temporarily visit other patches. The flux and simple trip models on two patches are expressed as Petri nets in Figure~\ref{fig:palette}(b.iii, b.iv). In~\cite{citron2021} the adjustments to the disease models are done manually and do not express a formal relationship between the adjusted models and their component disease and movement models. By contrast, our approach to model stratification formalizes this construction. In particular, the stratification of the Petri nets for the SIR and SIS disease models (Figure~\ref{fig:type_systems}(b) and Figure~\ref{fig:palette}(a.i)) by the Petri nets for the flux and simple trip movement models mirror the differential equation models defined in~\cite[Equations 6, 7, 9, 10]{citron2021}. Our software implementation of the mathematical ideas presented in this section can then be applied to automatically generate the stratified models from the palette of component models. As shown in the Supplementary Material, this implementation greatly reduces the size of the Petri nets that must be encoded by hand.  Additionally, this approach makes it easier to extend the methods of~\cite{citron2021}, since new stratification schemes, once defined and typed, can be seamlessly integrated into the model construction, calibration, and analysis pipeline, instead of requiring experts to manually adjust each candidate disease model by each candidate movement model.

Mathematically, a stratified model is a \textit{pullback} of whole-grain Petri nets, or equivalently a \textit{product} in the slice category over the given type system. Properties of these well-studied categorical formalisms immediately verify useful properties of stratified Petri nets. For example, consider stratifying a disease model by quarantine status and by spatial dynamics. Since pullback is an associative and commutative binary operation, the order of stratification does not affect the final model. That is, the following procedures are equivalent: stratifying the disease model by spatial dynamics and then by quarantine status, stratifying the disease model by quarantine status and then by spatial dynamics, and stratifying the disease model by the stratification of spatial dynamics by quarantine status (or vice versa).


\section{Calibrating and analyzing models} \label{sec:analysis}

The purpose of epidemiological modeling during an unfolding pandemic is to transform sparse data into effective policy decisions. The robustness of this process depends on understanding how adjusting a model affects its accuracy in representing the data (model calibration) and the policy outcomes it evidences (model analysis). In Section~\ref{sec:compositional}, we described the mathematics and implementation of specifying models by composing open Petri nets and other differential equation models. In this section, we show how this modeling framework streamlines the iterative loop of model specification, calibration, and analysis in the context of composing Petri nets, clarifying several of the advantages sketched in Section~\ref{sec:advantages}.

Since our approach decouples disease models based on Petri nets from the implementation of simulators in code, analysis tools can be defined directly on the combinatorial data structures representing the Petri nets. A tool defined once can thus be applied with equal ease to explicitly defined Petri nets, compositions of Petri nets, hierarchically defined Petri nets, and stratified Petri nets. One class of analysis tools comes from integrating AlgebraicPetri with the SciML suite, which provides procedures for parameter estimation and sensitivity analysis. We illustrate how this integration tightens the modeling workflow using the example of the SVIIvR Petri net model defined in Figure~\ref{fig:sviivr}.


AlgebraicPetri includes a method that converts generic Petri nets into the reaction network format supported by Catalyst, a library in the SciML ecosystem which uses symbolic algebra from the ModelingToolkit framework to represent chemical reaction networks~\cite{ma2021modelingtoolkit}. This method can be applied independently of how the Petri net was constructed. For example, we can apply it to the modularly constructed SVIIvR model from Section~\ref{sec:petrinets} and use Catalyst to fit the parameters according to COVID-19 infection data gathered from the US state Georgia over a five month period~\cite{nytimes_covid}. In this case, there is an order of magnitude difference between the estimated initial population and the true initial population. This mismatch may trigger an adjustment to the underlying model such as by a parallel refinement of one or more of the submodels. The adjusted models can be fed into the same calibration pipeline with no modifications to the analysis code.



\begin{figure}[htbp]
    \centering
    \begin{subfigure}[b]{0.49\textwidth}
        \includegraphics[width=1\textwidth]{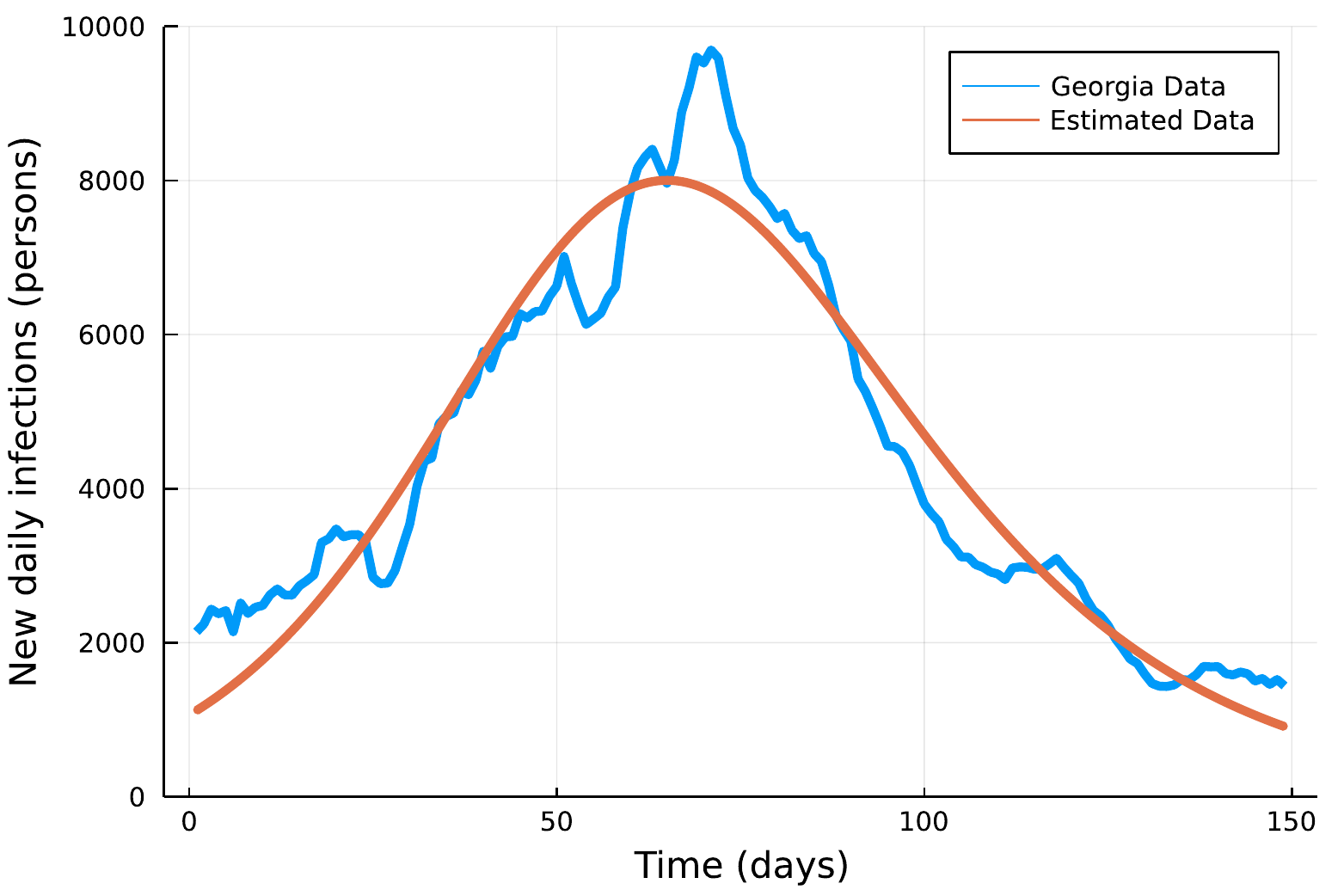}
        \caption{Result from fitting to Georgia infection data}
        \label{fig:calibraion_fit}
    \end{subfigure}
    \hspace{0.02\textwidth}
    \begin{subfigure}[b]{0.47\textwidth}
        \includegraphics[width=1\textwidth]{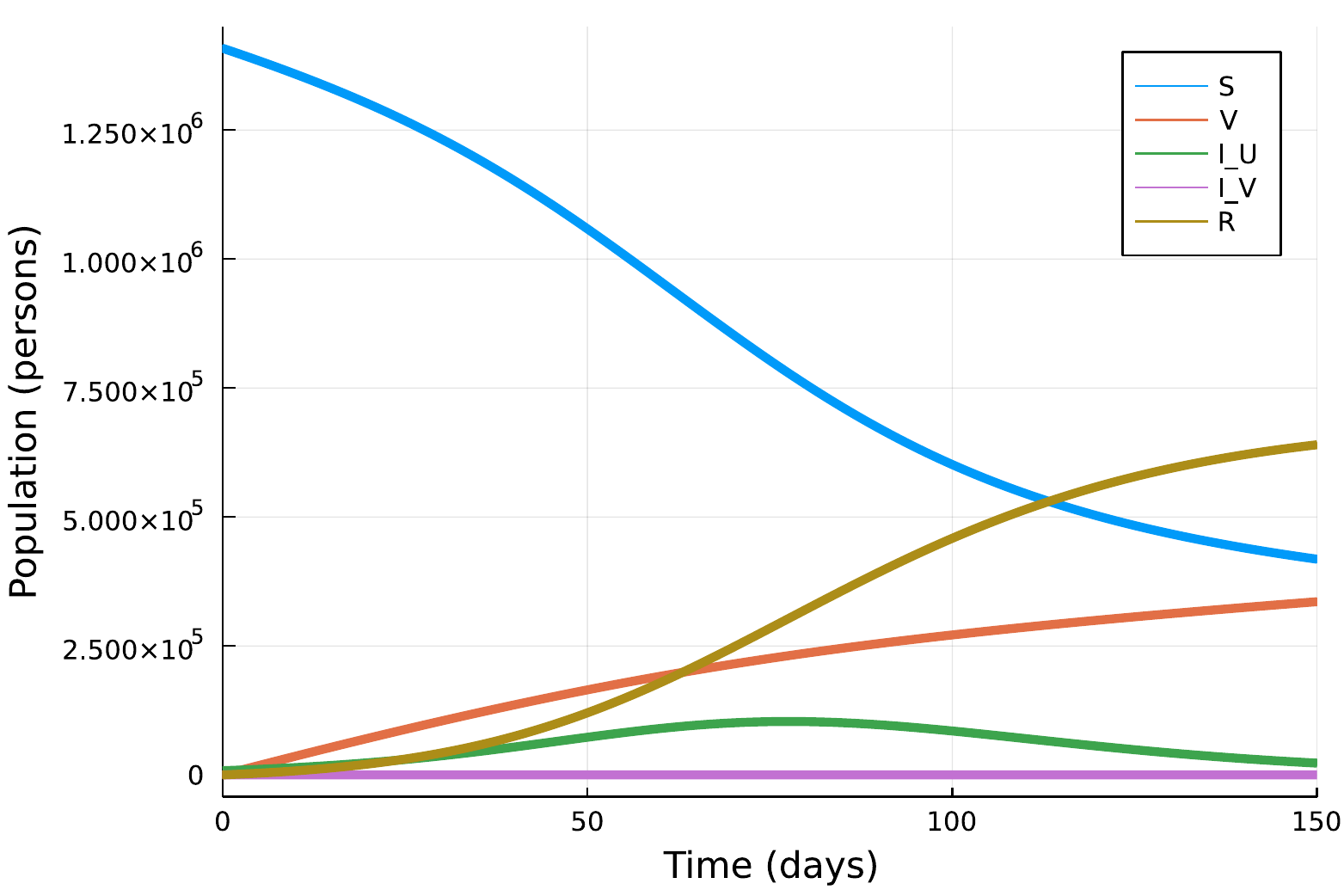}
        \caption{Petri net trajectory using the fit parameters}
        \label{fig:calibration_sim}
    \end{subfigure}
    \caption{Plots demonstrating the parameter estimation provided by integration with Catalyst. This example assumes a universal recovery rate of $\frac{1}{14}$ and an initial vaccinated population of zero. The remaining rates and initial populations are estimated.}
    \label{fig:calibration}
\end{figure}

\begin{figure}[htbp]
    \centering
    \begin{subfigure}[b]{0.45\textwidth}
    \begin{tabular}{p{0.3\textwidth}p{0.5\textwidth}}
    \toprule
    Transition & Sensitivity\\
    \midrule
    $\gamma_U$ & $5.27 \times 10^{-2}$\\
    $\nu$ & $4.29 \times 10^{-3}$\\
    $\gamma_V$ & $2.10 \times 10^{-6}$\\
    $\beta_{VV}$ & $-2.18 \times 10^{-10}$\\
    $\beta_{UV}$ & $-5.89 \times 10^{-7}$\\
    $\beta_{VU}$ & $-4.61 \times 10^{-6}$\\
    $\beta_{UU}$ & $-8.49 \times 10^{-2}$\\
    \bottomrule
    \end{tabular}
    \caption{Table of transition sensitivities}
    \label{tab:sensitivity}
    \end{subfigure}
    \begin{subfigure}[b]{0.45\textwidth}
    \includegraphics[width=1\textwidth]{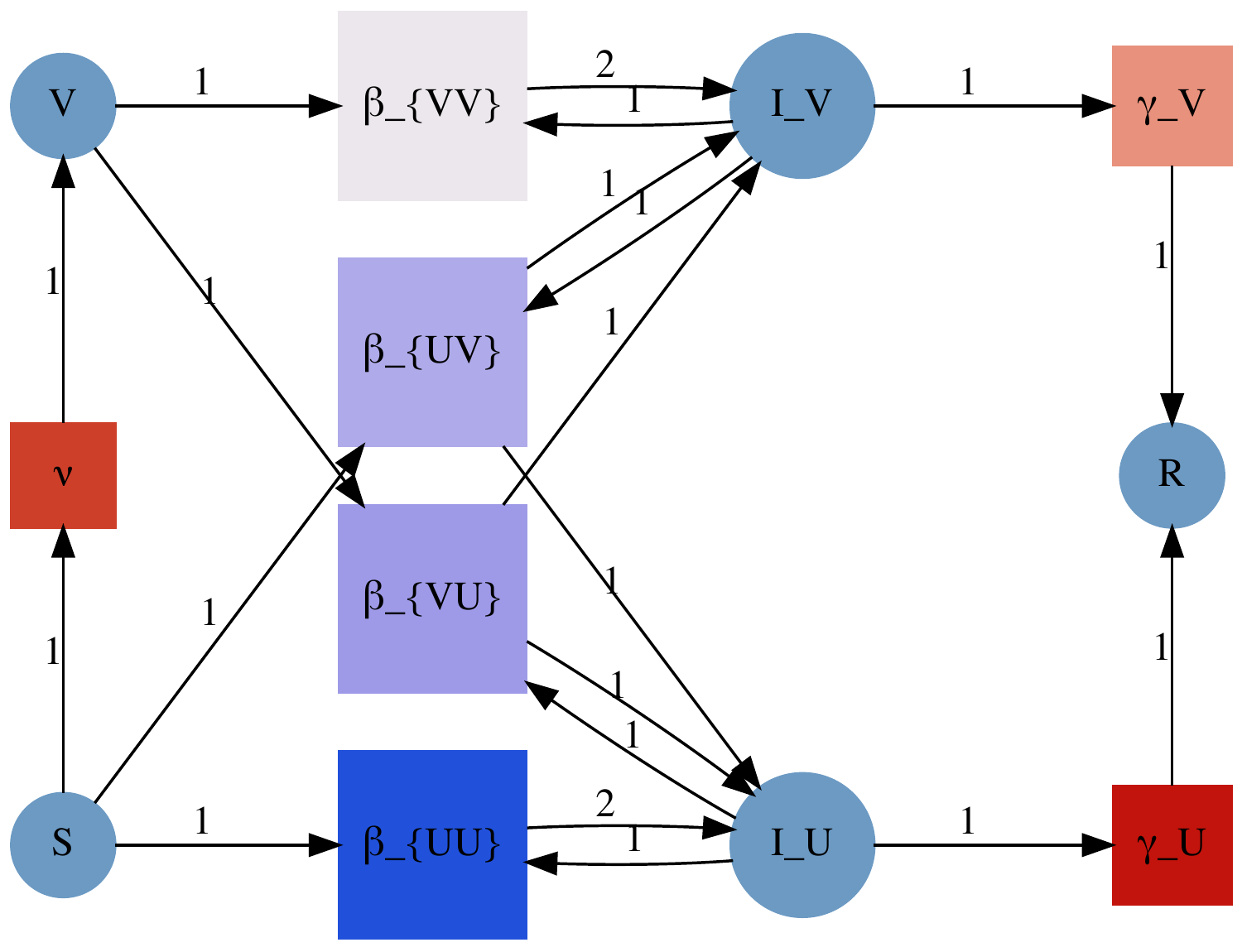}
    \caption{Petri net heatmap of sensitivities}
    \label{fig:sensitivity_heatmap}
    \end{subfigure}
    \caption{ In this example, the model outcome is the integral of the non-infectious populations over the course of the pandemic, and the sensitivities are measured for the parameters fit in Figure \ref{fig:calibration}. The sensitivities are tabulated in (a) and visualized as a heatmap in (b). In the Petri net heatmap, red and blue colors correspond to positive and negative sensitivities, respectively. }
    \label{fig:sensitivity}
\end{figure}

Following calibration, a model can be analyzed to suggest policy decisions and predict policy outcomes. A seamless process for analyzing models as they evolve is critical to test the robustness of these decisions and outcomes.
As an example analysis, we integrate the proportion of the non-infectious population over a simulation of the SVIIvR model. We also compute the sensitivity of this outcome to the transition rates of the Petri net using the tools in ForwardDiff~\cite{RevelsLubinPapamarkou2016}. In Figure~\ref{fig:sensitivity}, the sensitivity results are visualized by a heatmap which explicitly connects the results of the analysis and the underlying Petri net model. Adjustments to the underlying model---such as adding or removing transitions, changing transition rates, or substituting one submodel for another---are immediately reflected in the analysis. This tight feedback loop gives practical and visual tools for determining which policy decisions or outcomes are robust to model changes. In these examples, the analysis is treated externally to the model with the analysis being run directly on the ODE derived from the Petri net. However, in future work we intend to incorporate explicit representations of behavioral analyses into the compositional framework and those which are informed by the structure of composition. These analyses may be purely observational, actively control submodels in the composite, or check satisfaction of contracts as formalized in~\cite{bakirtzis2021cyberphysical}.

\section{Conclusion}

Scientific modeling is an iterative process of proposing, implementing, simulating, calibrating, analyzing, and comparing models. We presented a mathematical framework and software tools to accelerate the modeling process for compartmental models of infectious disease, in an effort to reduce the response time to emerging pandemics. Our framework is grounded in applied category theory and captures the algebraic and compositional structure of scientific models in a way that can be easily conveyed to both human scientists and computer systems. As a result, complex models can be specified compositionally using the syntax of wiring diagrams and algebraic operations or through the stratification of typed models. 
Our approach makes model structure a readily computable resource, which streamlines numerous downstream analyses, such as parameter estimation and sensitivity analysis. Together, the mathematical and computational features of our approach simplify and accelerate the iterative modeling process.

The structuralist approach to epidemiological modeling suggests many directions for future work. It can be extended to incorporate additional model semantics, such as stock-and-flow diagrams as an alternative to Petri nets, or stochastic and jump differential equations as complements to ODEs and DDEs. As demonstrated, the compositional structure simplifies the specification and visibility of multi-faceted models. A natural next step would be to investigate how compositional structure can be exploited in the mathematical and computational analysis of the models. For instance, parallel computations could be organized using the hierarchical decomposition already inherent in the model specification.\\

\noindent {\bf Funding} The authors were supported by the following DARPA Awards W911NF2110323 (Fairbanks), HR00112090067 (Libkind and Patterson), and HR00111990008 (Baas and Halter) along with AFOSR Award FA9550-20-1-0348 (Patterson).\\

\noindent {\bf Acknowledgements} The authors thank Xiaoyan Li, Nathaniel Osgood, David Smith, and Sean Wu for valuable insights into the methods and workflows of professional epidemiological modelers. We are also grateful to John Baez for insights into categories and pullbacks of Petri nets. We thank Alexandra Trani and Sean Wu for thorough reviews of the manuscript.

\bibliographystyle{plain}
\bibliography{references}

\begin{thebibliography}{10}

\bibitem{baez2020structured}
John~C. Baez and Kenny Courser.
\newblock Structured cospans.
\newblock {\em Theory and Applications of Categories}, 35(48):1771--1822, 2020.

\bibitem{baezOpenPetriNets2020}
John~C. Baez and Jade Master.
\newblock Open {{Petri}} nets.
\newblock {\em Mathematical Structures in Computer Science}, 30(3):314--341,
  2020.

\bibitem{baezCompositionalFrameworkReaction2017}
John~C. Baez and Blake~S. Pollard.
\newblock A {{Compositional Framework}} for {{Reaction Networks}}.
\newblock {\em Reviews in Mathematical Physics}, 29(09):1750028, 2017.

\bibitem{bakirtzis2021cyberphysical}
Georgios Bakirtzis, Cody~H. Fleming, and Christina Vasilakopoulou.
\newblock Categorical semantics of cyber-physical systems theory.
\newblock {\em ACM Transactions on Cyber-Physical Systems}, 5(3):1–32, Jul
  2021.

\bibitem{coexist}
Gergo Bohner, Gaurav Venkataraman, and Harrison Wilde.
\newblock {COEXI(S)T}: Modelling {COVID}-19 exit strategies for policy makers
  in the {United Kingdom}.
\newblock \url{https://github.com/gbohner/coexist/}, 2020.

\bibitem{kappa}
P.~Boutillier, J.~Feret, J.~Krivine, and W.~Fontana.
\newblock The {Kappa} language and tools.
\newblock \url{https://kappalanguage.org/}, 2021.

\bibitem{stan}
Bob Carpenter, Andrew Gelman, Matthew~D. Hoffman, Daniel Lee, Ben Goodrich,
  Michael Betancourt, Marcus Brubaker, Jiqiang Guo, Peter Li, and Allen
  Riddell.
\newblock Stan: A probabilistic programming language.
\newblock {\em Journal of statistical software}, 76(1):1--32, 2017.

\bibitem{citron2021}
Daniel~T. Citron, Carlos~A. Guerra, Andrew~J. Dolgert, Sean~L. Wu, John~M.
  Henry, David~L. Smith, et~al.
\newblock Comparing metapopulation dynamics of infectious diseases under
  different models of human movement.
\newblock {\em Proceedings of the National Academy of Sciences}, 118(18), 2021.

\bibitem{fong2015decorated}
Brendan Fong.
\newblock Decorated cospans.
\newblock {\em Theory and Applications of Categories}, 30(33):1096--1120, 2015.

\bibitem{friedmanPredictivePerformanceInternational2021}
Joseph Friedman, Patrick Liu, Christopher~E. Troeger, Austin Carter, Robert~C.
  Reiner, Ryan~M. Barber, James Collins, Stephen~S. Lim, David~M. Pigott, Theo
  Vos, Simon~I. Hay, Christopher J.~L. Murray, and Emmanuela Gakidou.
\newblock Predictive performance of international {{COVID}}-19 mortality
  forecasting models.
\newblock {\em Nature Communications}, 12(1):2609, 2021.

\bibitem{grimm2010odd}
Volker Grimm, Uta Berger, Donald~L. DeAngelis, J.~Gary Polhill, Jarl Giske, and
  Steven~F. Railsback.
\newblock The {{ODD}} protocol: {A} review and first update.
\newblock {\em Ecological Modelling}, 221(23):2760–2768, Nov 2010.

\bibitem{copasi}
Stefan Hoops, Sven Sahle, Ralph Gauges, Christine Lee, J{\"u}rgen Pahle,
  Natalia Simus, Mudita Singhal, Liang Xu, Pedro Mendes, and Ursula Kummer.
\newblock {COPASI}: a complex pathway simulator.
\newblock {\em Bioinformatics}, 22(24):3067--3074, 2006.

\bibitem{sbml}
Michael Hucka, Andrew Finney, Herbert~M. Sauro, Hamid Bolouri, John~C. Doyle,
  Hiroaki Kitano, Adam~P. Arkin, Benjamin~J. Bornstein, Dennis Bray, Athel
  Cornish-Bowden, et~al.
\newblock The {Systems Biology Markup Language} ({SBML}): a medium for
  representation and exchange of biochemical network models.
\newblock {\em Bioinformatics}, 19(4):524--531, 2003.

\bibitem{king2016pomp}
Aaron~A. King, Dao Nguyen, and Edward~L. Ionides.
\newblock Statistical inference for partially observed {{Markov}} processes via
  the {{R}} package pomp.
\newblock {\em Journal of Statistical Software}, 69(12), 2016.

\bibitem{kock2020}
Joachim Kock.
\newblock Elements of {{Petri}} nets and processes.
\newblock In {\em Applied Category Theory 2020}, 2020.

\bibitem{libkind2021operadic}
Sophie Libkind, Andrew Baas, Evan Patterson, and James Fairbanks.
\newblock Operadic modeling of dynamical systems: {Mathematics} and
  computation.
\newblock In {\em Applied Category Theory 2021}, 2021.

\bibitem{ma2021modelingtoolkit}
Yingbo Ma, Shashi Gowda, Ranjan Anantharaman, Chris Laughman, Viral Shah, and
  Chris Rackauckas.
\newblock {ModelingToolkit}: A composable graph transformation system for
  equation-based modeling, 2021.

\bibitem{RevelsLubinPapamarkou2016}
J.~{Revels}, M.~{Lubin}, and T.~{Papamarkou}.
\newblock Forward-mode automatic differentiation in {Julia}.
\newblock {\em arXiv:1607.07892}, 2016.

\bibitem{sharpe1978}
F.R. Sharpe and Alfred~J. Lotka.
\newblock Contribution to the analysis of malaria epidemiology. {IV}.
  {I}ncubation lag.
\newblock In {\em The Golden Age of Theoretical Ecology: 1923--1940}, pages
  348--368. Springer, 1978.

\bibitem{smith2014}
D.~L. Smith, T.~A. Perkins, R.~C. Reiner, C.~M. Barker, T.~Niu, L.~F. Chaves,
  A.~M. Ellis, D.~B. George, A.~Le~Menach, J.~R.~C. Pulliam, D.~Bisanzio,
  C.~Buckee, C.~Chiyaka, D.~A.~T. Cummings, A.~J. Garcia, M.~L. Gatton, P.~W.
  Gething, D.~M. Hartley, G.~Johnston, E.~Y. Klein, E.~Michael, A.~L. Lloyd,
  D.~M. Pigott, W.~K. Reisen, N.~Ruktanonchai, B.~K. Singh, J.~Stoller, A.~J.
  Tatem, U.~Kitron, H.~C.~J. Godfray, J.~M. Cohen, S.~I. Hay, and T.~W. Scott.
\newblock Recasting the theory of mosquito-borne pathogen transmission dynamics
  and control.
\newblock {\em Transactions of the Royal Society of Tropical Medicine and
  Hygiene}, 108(4):185--197, 2014.

\bibitem{spivak2013}
David~I. Spivak.
\newblock The operad of wiring diagrams: formalizing a graphical language for
  databases, recursion, and plug-and-play circuits.
\newblock {\em arXiv:1305.0297}, 2013.

\bibitem{nytimes_covid}
The New~York Times.
\newblock Coronavirus ({{COVID}}-19) data in the {{United States}}.
\newblock \url{https://github.com/nytimes/covid-19-data}, 2021.
\newblock Retrieved September 3, 2021.

\bibitem{vagner2015}
Dmitry Vagner, David~I. Spivak, and Eugene Lerman.
\newblock Algebras of open dynamical systems on the operad of wiring diagrams.
\newblock {\em Theory and Applications of Categories}, 30(51):1793–1822,
  2015.

\bibitem{netlogo}
U.~Wilensky.
\newblock Netlogo.
\newblock \url{http://ccl.northwestern.edu/netlogo/}, 1999.

\bibitem{wu2021}
Sean~L. Wu, Jared~B. Bennett, H{\'e}ctor~M. S{\'a}nchez~C., Andrew~J. Dolgert,
  Tom{\'a}s~M. Le{\'o}n, and John~M. Marshall.
\newblock {{MGDrivE}} 2: {{A}} simulation framework for gene drive systems
  incorporating seasonality and epidemiological dynamics.
\newblock {\em PLOS Computational Biology}, 17(5):e1009030, May 2021.

\end{thebibliography}

\section*{Supplementary Material}
We include the output of two Jupyter notebooks which reproduce the examples in this paper. The first notebook \textit{Compositional methods of model specification} implements the examples from Section~\ref{sec:compositional}. The second notebook \textit{Type systems for open Petri nets} implements the examples from Section~\ref{sec:stratification}. These notebooks along with the code for the calibration and analysis pipeline discussed in Section~\ref{sec:analysis} are available on GitHub at \url{https://github.com/AlgebraicJulia/Structured-Epidemic-Modeling/}.



\section*{Supplementary material}

```
In [2]:  using Catlab, Catlab.CategoricalAlgebra, Catlab.Programs, Catlab.WiringDiagrams, Catlab.Graphics
         using AlgebraicPetri
         using AlgebraicDynamics.UWDDynam
         using LabelledArrays
         using OrdinaryDiffEq, DelayDiffEq
         using Plots
```

# Compositional methods of model specification

This notebook develops compositional models in the style of structured multicospans and undirected wiring diagrams. It reproduces the examples in Section 2 of *An Algebraic Framework for Structured Epidemic Modeling*.

## Structured Multicospans of Petri nets

In this section we specify a SVIIvR disease model as the composition of three submodels:

- An SIR model for unvaccinated individuals.
- An SIR model for vaccinated individuals (which we call the VIvR model).
- A cross-exposure model which dictates the interactions between unvaccinated and vaccinated individuals.

This construction follows the figure:

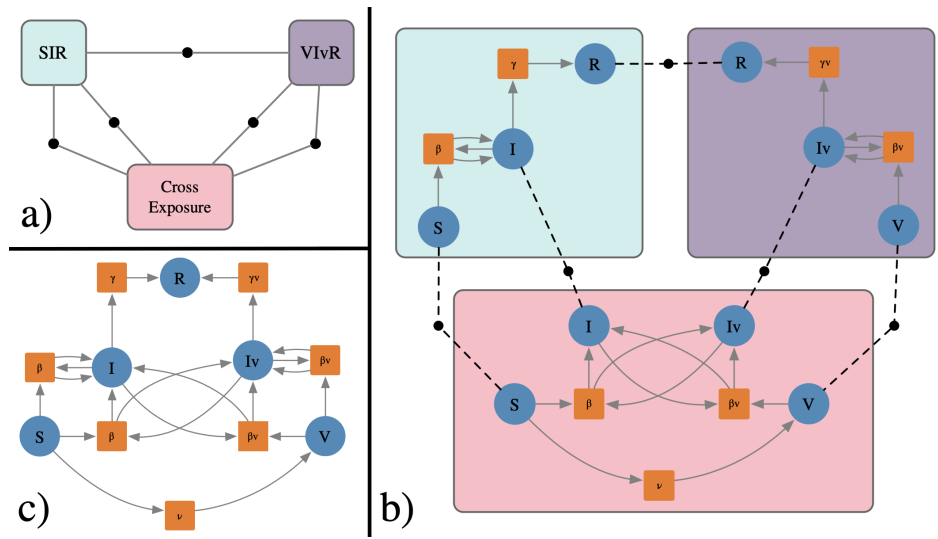

First we specify the composition pattern which dictates the interactions between the three submodels. In particular the SIR and VIvR models share the recovered population (R), the SIR and cross-exposure models share the susceptible and infected unvaccinated populations (S, I), and the VIvR and cross-exposure models share the susceptible and infected vaccinated populations (V, Iv).

```
In [3]:  SVIIvR_composition_pattern = @relation (S, V, I, Iv, R) where (S, V, I, Iv, R) begin
             SIR(S, I, R)
             VIvR(V, Iv, R)
             cross_exposure(S, I, V, Iv)
         end

         to_graphviz(SVIIvR_composition_pattern,
             box_labels = :name, junction_labels = :variable, edge_attrs=Dict(:len => "1"))
```

Out[3]:

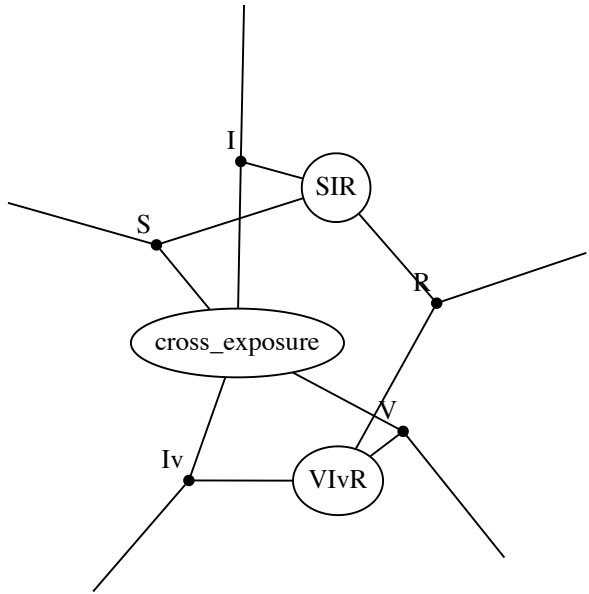

Next we define the Petri nets for the component submodels.

In [4]:
```
SIR = Open(LabelledPetriNet([:S, :I, :R],
    :inf_uu => ((:S, :I) => (:I, :I)),
    :rec_u => (:I => :R)
))

Graph(SIR)
```

Out[4]:
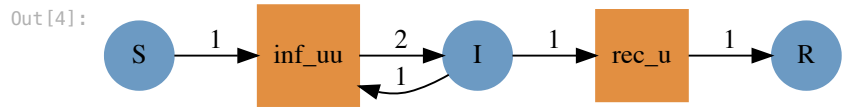

In [5]:
```
VIvR = Open(LabelledPetriNet([:V, :Iv, :R],
    :inf_vv => ((:V, :Iv) => (:Iv, :Iv)),
    :rec_v => (:Iv => :R)
))

Graph(VIvR)
```

Out[5]:
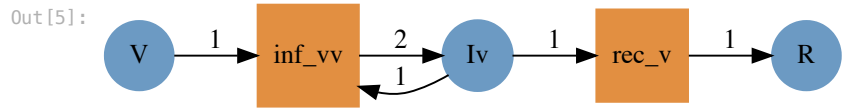

In [6]:
```
cross_exposure = Open(LabelledPetriNet([:S, :I, :V, :Iv],
    :inf_uv => ((:S, :Iv) => (:I, :Iv)),
    :inf_vu => ((:V, :I) => (:Iv, :I)),
    :vax => (:S => :V)
))

Graph(cross_exposure)
```

Out[6]:

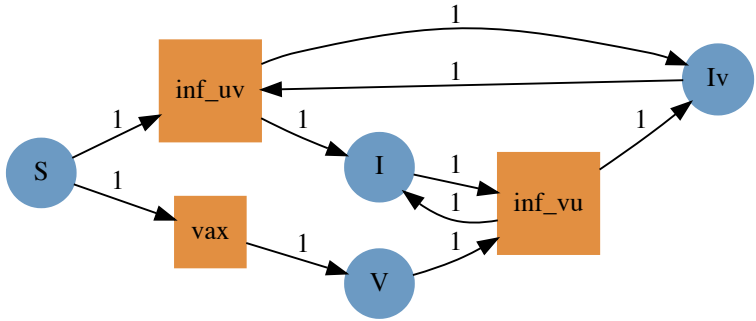

Finally, we compose the component submodels according to the interaction pattern defined by `SVIIvR_composition_pattern` using the `oapply` method. The result is the SVIIvR model of disease dynamics which accounts for vaccination.

In [7]:
```
SVIIvR = oapply(SVIIvR_composition_pattern, Dict(
    :SIR => SIR,
    :VIvR => VIvR,
    :cross_exposure => cross_exposure
)) |> apex

Graph(SVIIvR)
```

Out[7]:

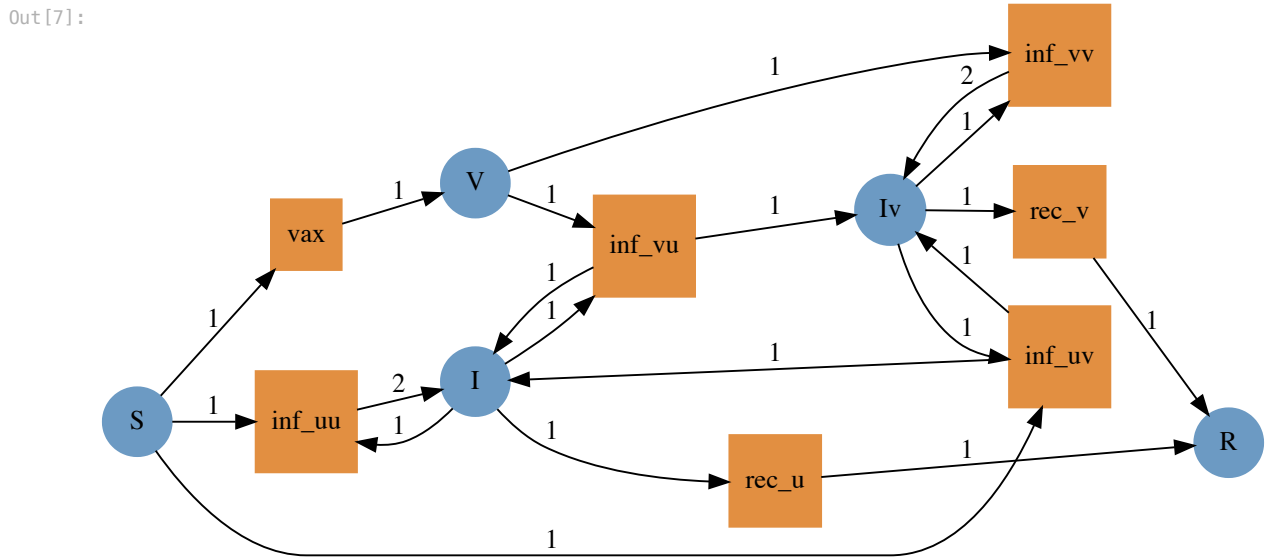

We additionally calibrate the parameters of this model with Catalyst.jl in the following: https://github.com/AlgebraicJulia/Structured-Epidemic-Modeling/tree/main/param_est.

## Composite disease model for vector borne diseases

In this section, we demonstrate composition of general differential equation models for vector-borne diseases. This composite has three components: disease dynamics in the host population, disease dynamics in the vector population, and the bloodmeal.

As in the previous example, we begin by specifying the composition pattern which specifies the interaction between the three components. In particular, the host dynamics and the bloodmeal share the infected host population (Ih), while the vector dynamics and the bloodmeal share the infected vector population (Iv). We visualize this composition pattern below. Notice that the disease dynamics in hosts and the disease dynamics in vectors only relate through the bloodmeal.

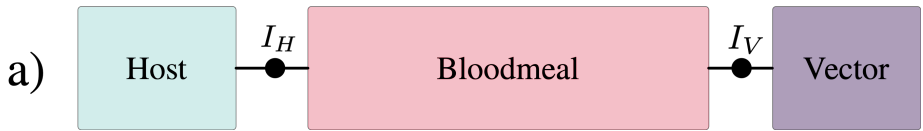

a)

```
In [8]:  bloodmeal_composition_pattern = @relation (Ih, Iv) where (Ih, Iv) begin
             host(Ih)
             bloodmeal(Ih, Iv)
             vector(Iv)
         end

         to_graphviz(bloodmeal_composition_pattern,  box_labels = :name, junction_labels = :variable)
```

Out[8]:

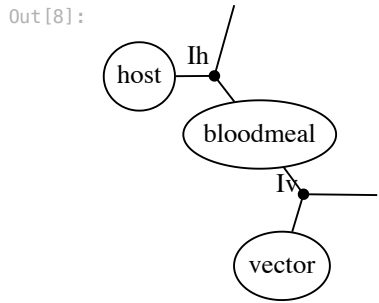

Next, we define the component models as ODEs. For each of the disease dynamics in hosts and vectors we specify a single model which represents the recovery of infected individuals at rates $r$ and $g$ respsectively.

```
In [9]:  host_dynamics = ContinuousResourceSharer{Float64}(1, 1, (u,p,t) -> -p.r*u, [1]);
         vector_dynamics = ContinuousResourceSharer{Float64}(1, 1, (u,p,t) -> -p.g*u, [1]);
```

For the bloodmeal, we define three model choices for this component. The first bloodmeal model is the law of mass action applied to the Petri net in (b) below. (c) gives the ODE interpretation of the Petri net.

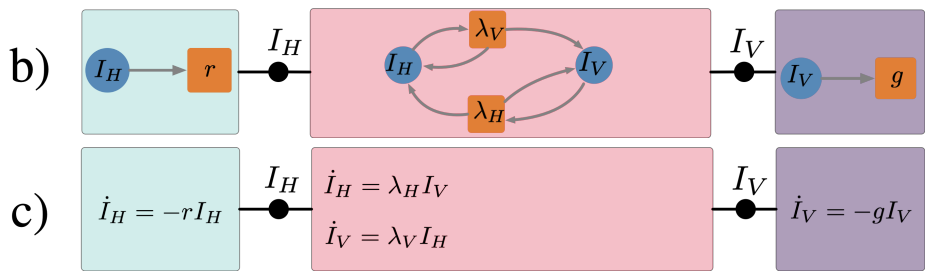

This model is given by the ODE

$$\dot{I}_H = \lambda_H I_V, \quad \dot{I}_V = \lambda_V I_H.$$

So the density of infected hosts grows at a rate $\lambda_H$ proportional to the density of infected vectors. Likewise, the density of infected vectors grows at a rate $\lambda_V$ proportional to the density of infected hosts.

```
In [10]:  bloodmeal_mass_action = ContinuousResourceSharer{Float64}(2, 2, (u,p,t) -> [p.λₕ*u[2], p.λₕ*u[1]], [1,2])
```

The second bloodmeal model is a Ross-Macdonald model given by the ODE

$$\dot{I}_H = \frac{1}{H} ab I_V (H - I_H), \quad \dot{I}_V = ac \frac{I_H}{H}(V - I_V)$$

where $a$ is the biting rate, $b$ is the transmission efficiency from infectious vectors to susceptible hosts, $c$ is the transmission efficiency from infectious hosts to susceptible vectors, $H$ is the total host population, and $V$ is the total vector population.

21

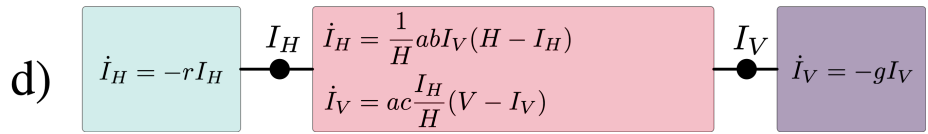

d)

```
In [11]:  function dudt(u,p,t)
            Ih, Iv = u
            return [(p.a*p.b/p.H)*Iv*(p.H - Ih), (p.a*p.c)*(Ih/p.H)*(p.V - Iv)]
          end
          bloodmeal_RossMacdonald = ContinuousResourceSharer{Float64}(2, 2, dudt, [1,2]);
```

In the third bloodmeal model, we extend the Ross-Macdonald model above by using delay differential equations to represent the incubation period for the disease in the mosquito population.

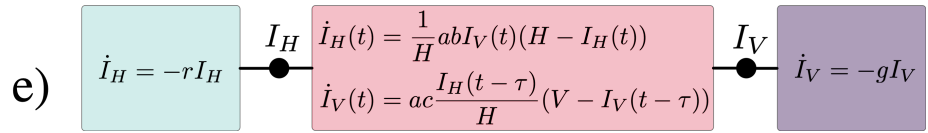

e)

```
In [12]:  function dudt_delay(u,h,p,t)
            Ih, Iv = u
            Ih_delay, Iv_delay = h(p, t - p.n)
            return [(p.a*p.b/p.H)*Iv*(p.H - Ih), (p.a*p.c)*(Ih_delay/p.H)*(p.V - Iv_delay)]
          end

          bloodmeal_delay = DelayResourceSharer{Float64}(2, 2, dudt_delay, [1,2]);

          host_delay   = DelayResourceSharer{Float64}(1, 1, (u,h,p,t) -> -p.r*u, [1]);
          vector_delay = DelayResourceSharer{Float64}(1, 1, (u,h,p,t) -> -p.g*u, [1]);
```

Lastly, we compose the component models according to the composition pattern `bloodmeal_composition_pattern` to specify three composite models for vector-borne disease dynamics.

```
In [13]:  malaria_mass_action = oapply(bloodmeal_composition_pattern, Dict(
            :host => host_dynamics,
            :vector => vector_dynamics,
            :bloodmeal => bloodmeal_mass_action
          ));

          malaria_RossMacdonald = oapply(bloodmeal_composition_pattern, Dict(
            :host => host_dynamics,
            :vector => vector_dynamics,
            :bloodmeal => bloodmeal_RossMacdonald
          ));

          malaria_delay = oapply(bloodmeal_composition_pattern, Dict(
            :host => host_delay,
            :vector => vector_delay,
            :bloodmeal => bloodmeal_delay
          ));
```

To illustrate the effect of substituting different choices for the bloodmeal component model, we solve and plot solutions for the composite models.

```
In [14]:  params = LVector(a = 0.3, b = 0.55, c = 0.15,
              g = 0.1, r = 1.0/200, V = 1000.0, H = 2000.0,
              n = 10.0,
              λₕ = 0., λᵥ = 0.)
          params[:λₕ] = params.a * params.b / params.H
          params[:λᵥ] = params.a * params.c / params.H

          u0 = [10.0, 100.0]
          tspan = (0., 365.);
```

```
In [15]:  prob = ODEProblem(malaria_mass_action, u0, tspan, params)
```

```julia
sol = solve(prob, Tsit5())
plot(sol,
    lw = 2,
    label = ["infected hosts" "infected vectors"],
    ylabel = "population density", xlabel = "time",
    title = "Mass Action Composite Model"
)
```

Out[15]:

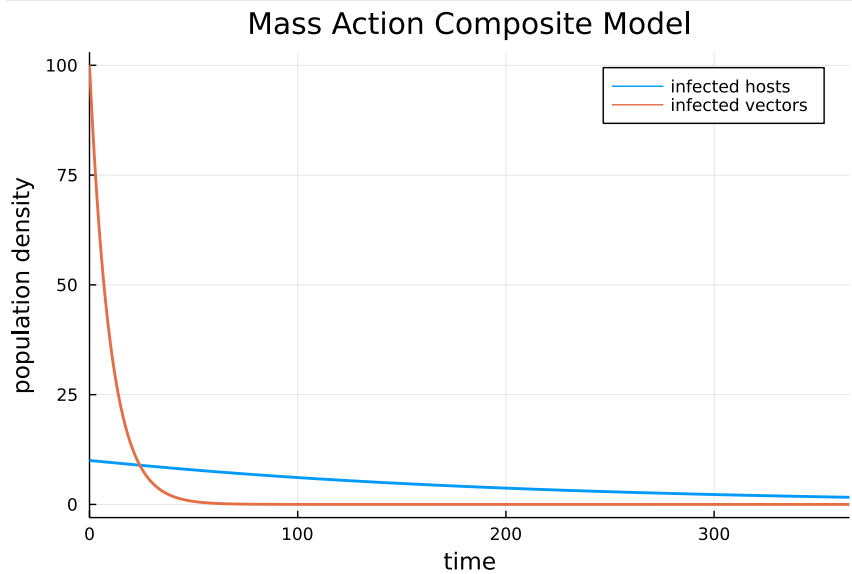

In [16]:
```julia
prob_ODE = ODEProblem(malaria_RossMacdonald, u0, tspan, params)
sol_ODE = solve(prob_ODE, Tsit5())

plot(sol_ODE,
    lw = 2,
    label = ["infected hosts - ODE model" "infected vectors - ODE model"],
    ylabel = "population density", xlabel = "time",
    title = "Ross-Macdonald Composite Models"
)


hist(p, t) = u0
alg = MethodOfSteps(Tsit5())
prob_DDE = DDEProblem(malaria_delay, u0, hist, tspan, params)
sol_DDE = solve(prob_DDE, alg,abstol=1e-12,reltol=1e-12)

plot!(sol_DDE,
    lw = 2,
    linestyle = :dash,
    label = ["infected hosts - DDE model" "infected vectors - DDE model"],
)
```

Out[16]:

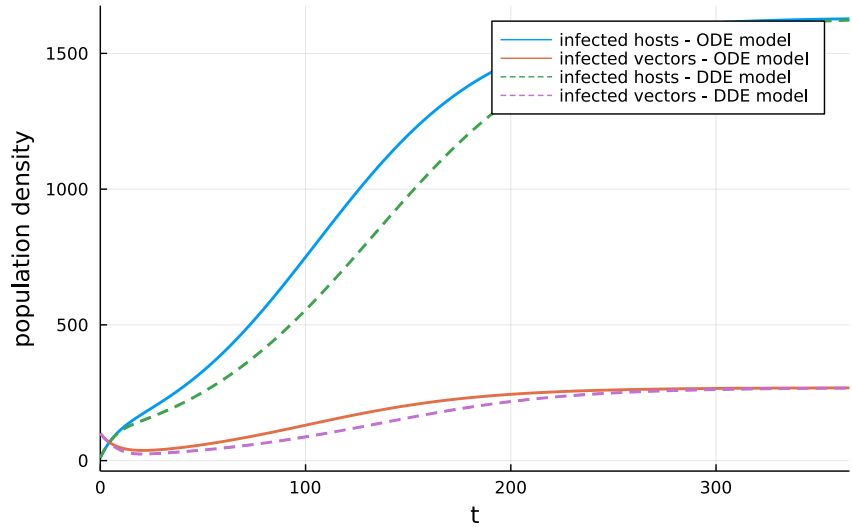

The plots demonstrate that the mass action bloodmeal model causes the density of infected hosts and infected vectors to vanish, whereas in the Ross Macdonald models the densities stabilize. These observations suggest that the Ross-Macdonald models are better suited to this modeling problem. On the other hand, the ordinary and delay differential equation Ross-Macdonald models reach the same equilibrium, although they exhibit different transient behavior before settling down.

This structured approach to model composition benefits the modeling workflow by providing tools to rapidly explore variations on model strucure in order to evaluate the cost-benefit tradeoffs for different models. In this case, we determined that the assumption of mass action is inconsistent with observed phenomena and that the ordinary and delay differential equation Ross-Macdonald models have equivalent equilibrium behavior. Modelers can thus consider their purpose in studying either transient or equilibrium behavior when deciding if the additional computational requirements of a DDE model are worth the improved fidelity to the transient phenomena.

This interactive model exploration environment where changes in a model's behavior can be tied back to changes in the model strucure, enables scientists to spend more time on their domain science and less time building, running, and analyzing software systems.

```julia
In [2]: using Catlab, Catlab.CategoricalAlgebra, Catlab.Programs, Catlab.WiringDiagrams, Catlab.Graphics.Graphvi
        using Catlab.Graphics.Graphviz: Html
        using AlgebraicPetri
```

```julia
In [3]: # Helper for graphing typed Petri nets
        colors = ["#a08eae","#ffeec6", "#a8dcd9", "#ffeec6", "#a8dcd9"]

        function def_trans(typed_petri::ACSetTransformation, colors; labels = true)
          (p, t; pos = "") -> ("t$t", Attributes(
                    :label => labels ? Html(flatten(tname(p,t))) : "" ,
                    :shape=>"square",
                    :color=>colors[typed_petri[:T](t)],
                    :pos=>pos))
        end

        function def_trans(colors = colors; labels = true)
          (p, t; pos = "") -> ("t$t", Attributes(
                    :label => labels ? "$(tname(p, t))" : "" ,
                    :shape=>"square",
                    :color=>colors[t],
                    :pos=>pos))
        end

        flatten(tname::Symbol) = "$tname"

        function flatten(tname::Tuple)
            names = split(replace(string(tname), "("=>"", ")"=>"", ":"=>""), ",")
            for i in 1:length(names)
                name = strip(names[i])
                if name[1:2] == "id"
                    continue
                end
                return name
            end
            return "id"
        end

        def_states(p, s; pos="") = ("s$s", Attributes(
                :label => sname(p,s) isa Tuple where T ? Html(replace(string(sname(p,s)), ":"=>"", "," => "<BR/>
                :shape=>"circle",
                :color=>"#6C9AC3",
                :pos=>pos
        ))

        Graph_typed(typed_petri::ACSetTransformation, colors = colors; labels = true) = Graph(dom(typed_petri),
            make_trans = def_trans(typed_petri, colors; labels = labels),
            make_states = def_states
        )
```

```
Out[3]: Graph_typed (generic function with 2 methods)
```

## Type systems for open Petri nets

This notebook showcases typed Petri nets and their application in generating stratified models. It reproduces the examples in Section 3 of *An Algebraic Framework for Structured Epidemic Modeling*.

### Typed Petri nets

We establish two Petri nets which are domain-specific type systems for epidemiological modeling.

$P_{\text{infectious}}$ represents a type system for infectious diseases. It has one place representing a single population and three transitions:

1. `interact` representing an interaction between two individuals.
2. `t_disease` representing a spontaneous change in disease status.
3. `t_strata` representing a spontaneous change in strata.

25

In [4]:
```
infectious_type = LabelledPetriNet([:Pop],
  :interact=>((:Pop, :Pop)=>(:Pop, :Pop)),
  :t_disease=>(:Pop=>:Pop),
  :t_strata=>(:Pop=>:Pop)
)

Graph_typed(id(infectious_type))
```

Out[4]:

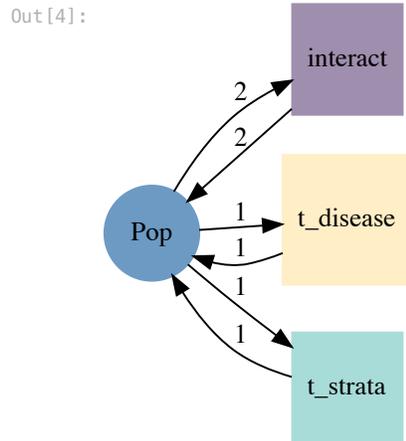

On the other hand, $P_{\text{vector-borne}}$ respresents a types sytem for vector-borne diseases. It has two places representing a host and a vector population. The transitions `t_disease` and `t_strata` represent spontaneous changes in disease and strata status for hosts and vectors. The single `interact` transition represents an interaction between a host and a vector.

In [5]:
```
vector_borne_type = LabelledPetriNet([:Host, :Vector],
  :interact=>((:Host, :Vector)=>(:Host, :Vector)),
  :t_disease=>(:Host=>:Host),
  :t_strata=>(:Host=>:Host),
  :t_disease=>(:Vector=>:Vector),
  :t_strata=>(:Vector=>:Vector)
)

Graph(vector_borne_type; make_trans = def_trans())
```

Out[5]:

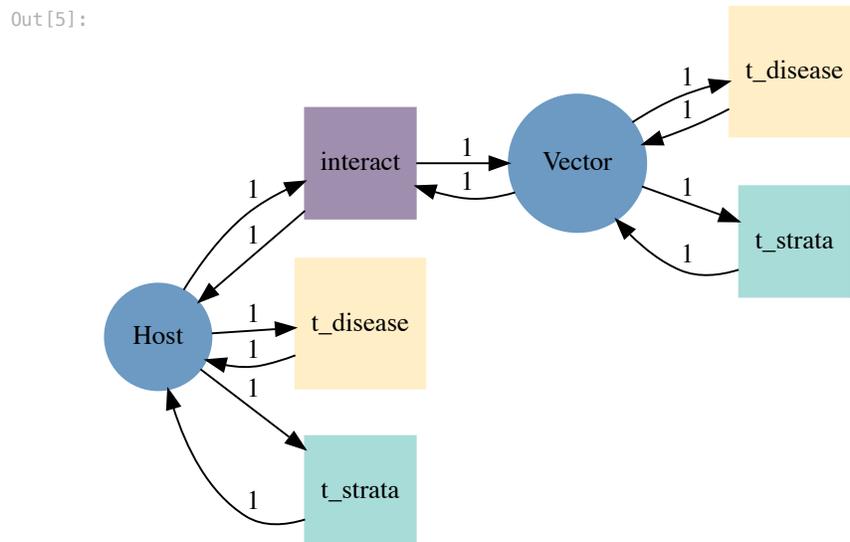

A Petri net typed by $P_{\text{infectious}}$ is a Petri net $P$ along with an etale map $P \to P_{\text{infectious}}$, which is captured by the data of an ACSetTransformation.

To make these ACSetTransformations easier to define, we begin by extracting data from the type system $P_{\text{infectious}}$.

26

In [6]:
```
s, = parts(infectious_type, :S)
t_interact, t_disease, t_strata = parts(infectious_type, :T)
i_interact1, i_interact2, i_disease, i_strata = parts(infectious_type, :I)
o_interact1, o_interact2, o_disease, o_strata = parts(infectious_type, :O);

infectious_type = map(infectious_type, Name=name->nothing); # remove names to allow for the loose ACSet
```

Our first example of a typed Petri net is the classic SIR model of disease infection. There are three places that are all typed by the single place type `Pop`, and five transitions. The transition labeled `inf` represents an infection between a susceptible and an infected individual. The transition labeled `rec` represents recovery of an infected individual. The transitions labeled `id` represent identity transformations or no change in disease status. However, they are important because in a stratification they may be paired with transitions that represent spontaneous change in strata.

In [7]:
```
SIR = LabelledPetriNet([:S, :I, :R],
    :inf => ((:S, :I)=>(:I, :I)),
    :rec => (:I=>:R),
    :id => (:S => :S),
    :id => (:I => :I),
    :id => (:R => :R)
)

typed_SIR = ACSetTransformation(SIR, infectious_type,
    S = [s, s, s],
    T = [t_interact, t_disease, t_strata, t_strata, t_strata],
    I = [i_interact1, i_interact2, i_disease, i_strata, i_strata, i_strata],
    O = [o_interact1, o_interact2, o_disease, o_strata, o_strata, o_strata],
    Name = name -> nothing # specify the mapping for the loose ACSet transform
);

@assert is_natural(typed_SIR)

Graph_typed(typed_SIR)
```

Out[7]:

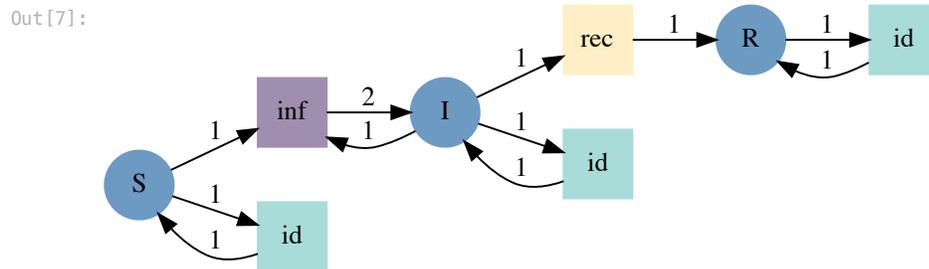

Our second example is of a stratification scheme for stratification by quarantine status. It contains a population of individuals in quarantine/isolation $Q$ and a population of individuals not in quarantine/isolation $Q^f$. Individuals can move between these strata and also undergo a spontaneous change in their infection status. However, interaction-type transformations can only occur between two individuals in the non-quarantining population.

In [8]:
```
quarantine = LabelledPetriNet([:Q, :NotQ],
    :id => (:Q => :Q),
    :id => (:NotQ => :NotQ),
    :enter_q => (:NotQ => :Q),
    :exit_q => (:Q => :NotQ),
    :interact => ((:NotQ, :NotQ) => (:NotQ, :NotQ))
)

typed_quarantine = ACSetTransformation(quarantine, infectious_type,
    S = [s, s],
    T = [t_disease, t_disease, t_strata, t_strata, t_interact],
    I = [i_disease, i_disease, i_strata, i_strata, i_interact1, i_interact2],
    O = [o_disease, o_disease, o_strata, o_strata, o_interact1, o_interact2],
    Name = name -> nothing
)

@assert is_natural(typed_quarantine)
```

```
Graph_typed(typed_quarantine)
```

Out[8]:

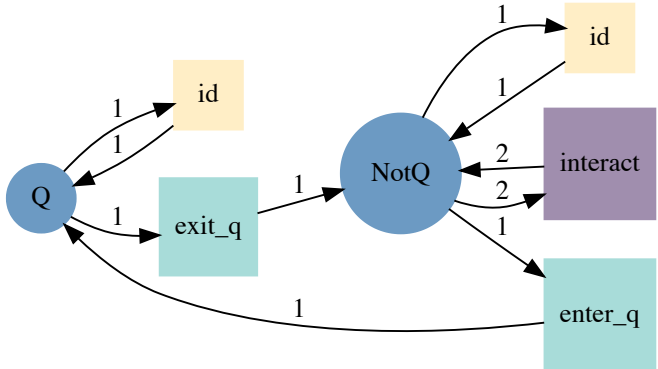

## Model stratification

This section presents examples of stratifying one typed model by another to build multi-faceted models.

### First example

First, we give methods to produce the stratified model of two typed Petri nets. `stratify` returns the stratified model while `typed_stratify` returns the stratified model along with a typing of it.

Mathematically, these operations correspond to pullbacks, and the `pullback` method is implemented by Catlab.jl. Graphically, you can see a pullback as shown in the following figure:

$$
\begin{array}{ccc}
P_{\text{stratified}} & \dashrightarrow & P_{\text{strata}} \\
\vdots & \lrcorner & \downarrow \\
\downarrow & & \downarrow \\
P_{\text{disease}} & \longrightarrow & P_{\text{infectious}}
\end{array}
$$

In [9]:
```
stratify(typed_model1, typed_model2) = ob(pullback(typed_model1, typed_model2))

typed_stratify(typed_model1, typed_model2) =
  compose(proj1(pullback(typed_model1, typed_model2)), typed_model1);
```

We apply this method to the SIR and quarantining models. Note that the labels of the places and transitions are pairs with the first element in the pair representing information from the disease model and the second element in the pair representing information from the stratification scheme. For example, the place labeled `(:S, :Q)` represents the population of susceptible individuals who are in quarantine. The transitions labeled `(:id, :exit_q)` represents no change (i.e. identity transformation) in disease status and a change in strata from quarantining to not quarantining. The transition labeled `(:inf, :interaction)` represents the infection interaction between two individuals who are not quarantining.

In [10]:
```
typed_stratified_model = typed_stratify(typed_SIR, typed_quarantine)
Graph_typed(typed_stratified_model)
```

Out[10]:

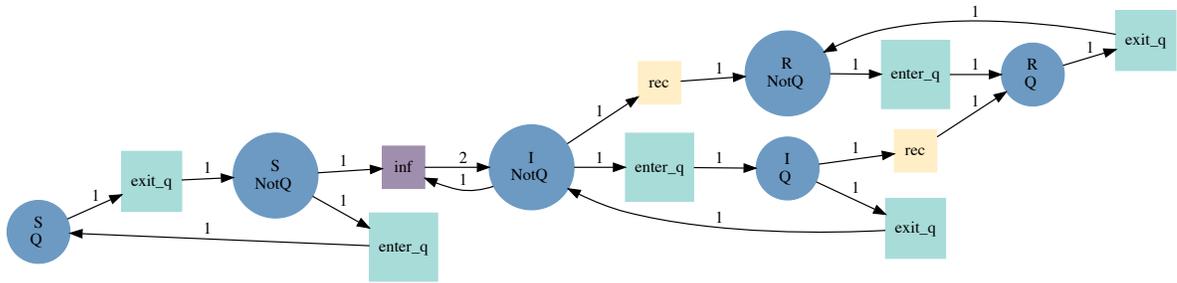

The layout engine used by Graphviz doesn't know about the pullback structure, so it doesn't generate a nice symmetric layout. Future work could explore algorithms for Petri net layout that use this structure. Below is the same Petri net but with a more systematic layout.

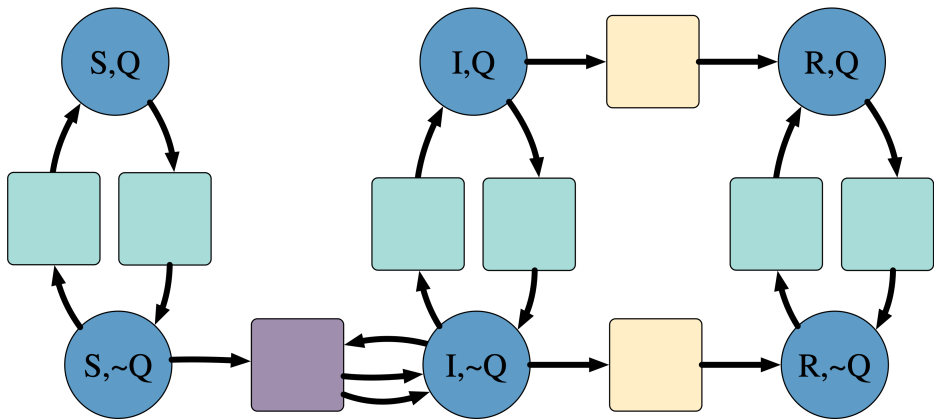

## Example playground

Next we give a palette of disease models and a palette of stratification schemes. These can be used to derive a number of stratified models as well as multiply-stratified models

### Palette of disease models

Our palette of disease models includes the SIR model implemented above along with the SIS and SVIIvR model pictured below:

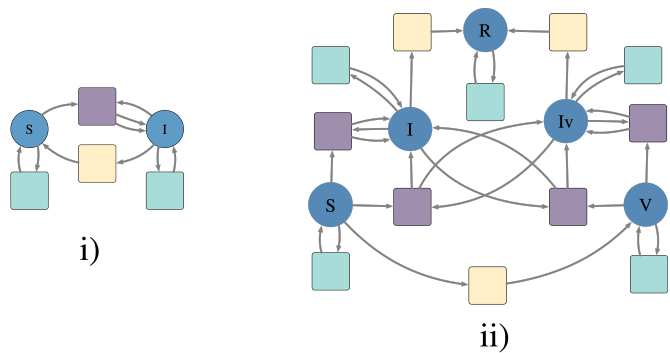

i)

ii)

Next, we describe and implement these additional disease models.

Our second disease model is the Susceptible-Infected-Susceptible (SIS) model of disease infection, in which there is no long-lasting immunity and a recovered indivdual is susceptible to reinfection.

In [11]:
```
SIS = LabelledPetriNet([:S, :I],
    :inf => ((:S, :I)=>(:I, :I)),
    :rec => (:I=>:S),
    :id => (:S => :S),
    :id => (:I => :I),
```

```
)
typed_SIS = ACSetTransformation(SIS, infectious_type,
  S = [s, s],
  T = [t_interact, t_disease, t_strata, t_strata],
  I = [i_interact1, i_interact2, i_disease, i_strata, i_strata],
  O = [o_interact1, o_interact2, o_disease, o_strata, o_strata],
  Name = name -> nothing
);

@assert is_natural(typed_SIS)

Graph_typed(typed_SIS)
```

Out[11]:

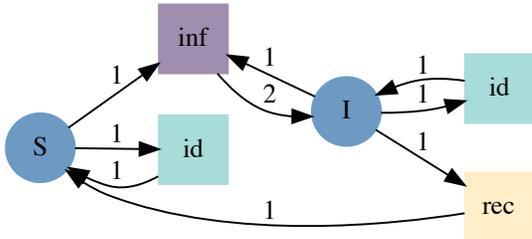

The third disease model is the SVIIvR model defined compositionally and discussed in detail in composition.ipynb. To this model we add identity transformations between each place that are typed by spontaneous change in strata.

In [12]:
```
SVIIvR_composition_pattern = @relation (S, V, I, Iv, R) where (S, V, I, Iv, R) begin
  SIR(S, I, R)
  VIvR(V, Iv, R)
  cross_exposure(S, I, V, Iv)
end

SIR = Open(LabelledPetriNet([:S, :I, :R],
  :inf_uu => ((:S, :I) => (:I, :I)),
  :rec_u => (:I => :R)
))

VIvR = Open(LabelledPetriNet([:V, :Iv, :R],
  :inf_vv => ((:V, :Iv) => (:Iv, :Iv)),
  :rec_v => (:Iv => :R)
))

cross_exposure = Open(LabelledPetriNet([:S, :I, :V, :Iv],
  :inf_uv => ((:S, :Iv) => (:I, :Iv)),
  :inf_vu => ((:V, :I) => (:Iv, :I)),
  :vax => (:S => :V)
))

SVIIvR = oapply(SVIIvR_composition_pattern, Dict(
  :SIR => SIR,
  :VIvR => VIvR,
  :cross_exposure => cross_exposure
)) |> apex;

ts = add_transitions!(SVIIvR, ns(SVIIvR), tname = :id)
add_inputs!(SVIIvR, ns(SVIIvR), ts, 1:ns(SVIIvR))
add_outputs!(SVIIvR, ns(SVIIvR), ts, 1:ns(SVIIvR));

typed_SVIIvR = ACSetTransformation(SVIIvR, infectious_type,
  S = [s, s, s, s, s],
  T = vcat([t_interact, t_disease, t_interact, t_disease, t_interact, t_interact, t_disease], repeat([t
  I = vcat([i_interact1, i_interact2, i_disease, i_interact1, i_interact2, i_disease, i_interact1, i_int
  O = vcat([o_interact1, o_interact2, o_disease, o_interact1, o_interact2, o_disease, o_interact1, o_int
  Name = name -> nothing
);

@assert is_natural(typed_SVIIvR)

Graph_typed(typed_SVIIvR)
```

Out[12]:

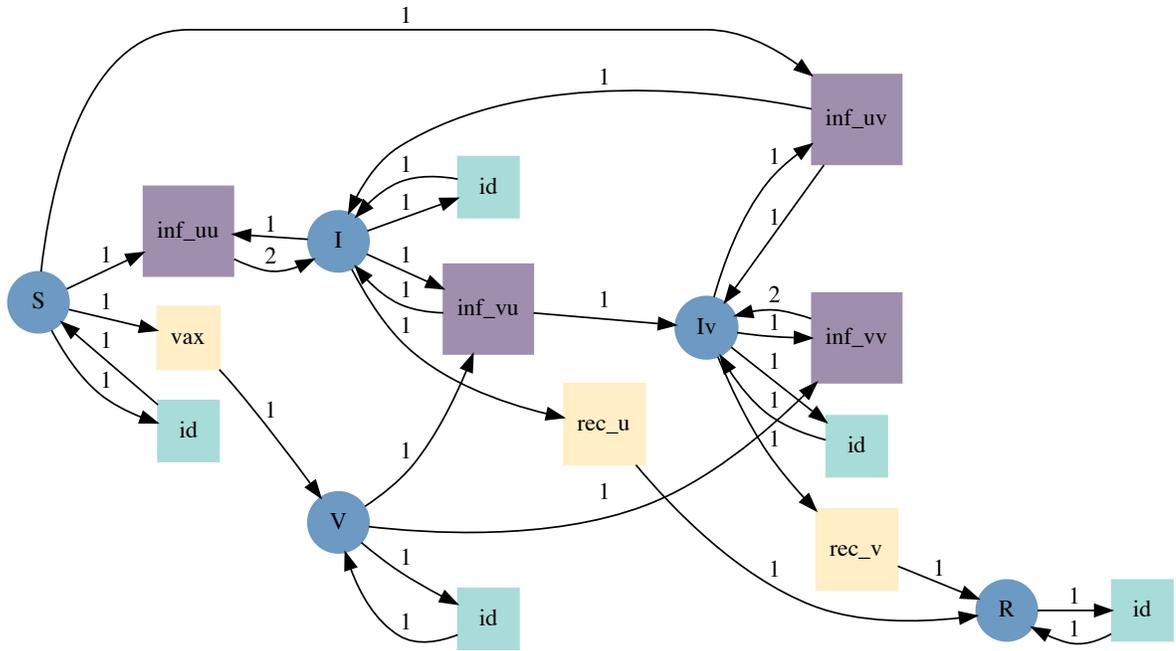

Finally, we define our palette of disease models.

In [13]:
```
disease_models = Dict(
  :SIR => typed_SIR,
  :SIS => typed_SIS,
  :SVIIvR => typed_SVIIvR,
);
```

We can stratify each of these disease models by the model of quarantining. Note that we omit the labels on the transition to reduce the size of the Petri net. Recall that blue transitions correspond to spontaneous changes in strata, yellow transitions correspond to spontaneous changes in disease status, and purple transitions correspond to interactions.

In [14]:
```
typed_model = typed_stratify(disease_models[:SIS], typed_quarantine)
Graph_typed(typed_model)
```

Out[14]:

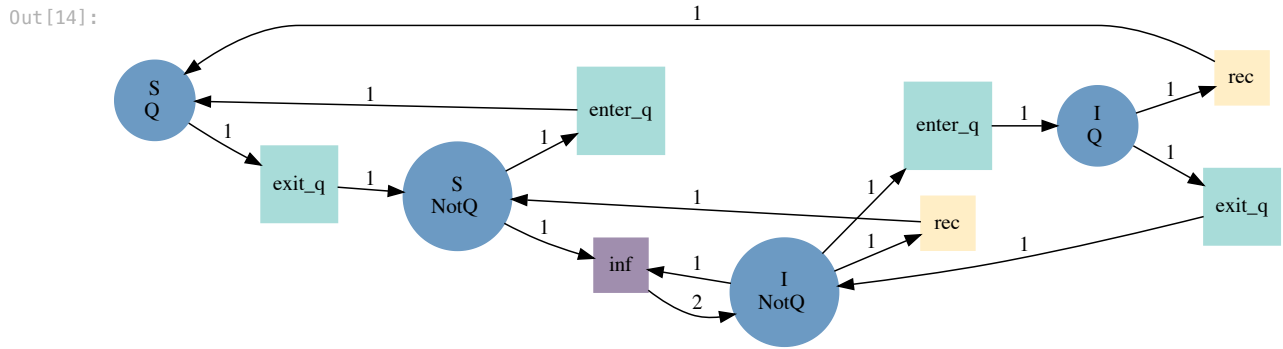

In [15]:
```
typed_model = typed_stratify(disease_models[:SVIIvR], typed_quarantine)
Graph_typed(typed_model)
```

Out[15]:

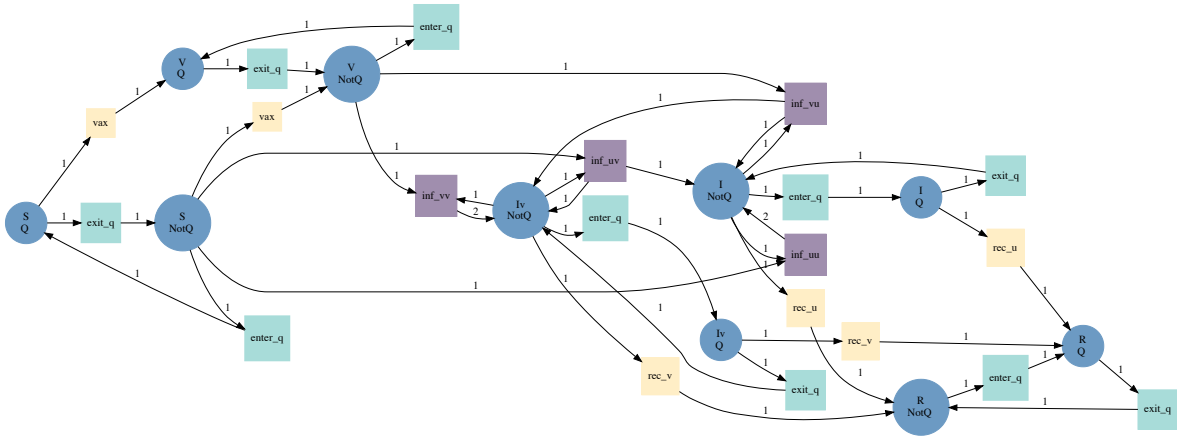

## Palette of stratification schemes

Our palette of stratification schemes includes the model of quarantine status (i) already implemented along with a model of age stratification (ii), and two movement models (iii and iv).

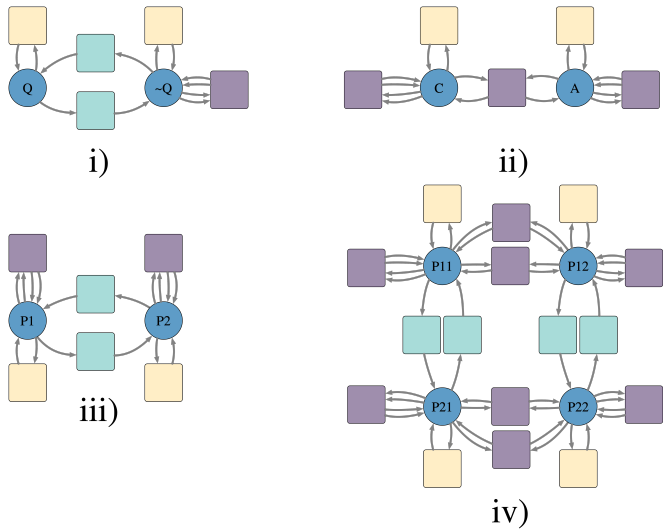

Next, we describe and implement these additional stratification schemes.

Our second stratification scheme is a stratification by age. In this model, individuals cannot move between strata. Individuals can undergo a spontaneous change in their infection status as well as interact with other individuals in their strata. Adults and children can also interact. However, the mapping of the input arcs asserts that children infect adults and not vice versa.

In [16]:
```
age_stratification = LabelledPetriNet([:Child, :Adult],
    :id => (:Child => :Child),
    :id => (:Adult => :Adult),
    :interact => ((:Child, :Child) => (:Child, :Child)),
    :interact => ((:Adult, :Adult) => (:Adult, :Adult)),
    :interact => ((:Child, :Adult) => (:Child, :Adult))
)

typed_age = ACSetTransformation(age_stratification, infectious_type,
    S = [s, s],
    T = [t_disease, t_disease, t_interact, t_interact, t_interact],
    I = [i_disease, i_disease, i_interact1, i_interact2, i_interact1, i_interact2, i_interact2, i_interact
    O = [o_disease, o_disease, o_interact1, o_interact2, o_interact1, o_interact2, o_interact2, o_interact
    Name = name -> nothing # specify the mapping for the loose ACSet transform
);
```

```
Graph_typed(typed_age)
```

Out[16]:

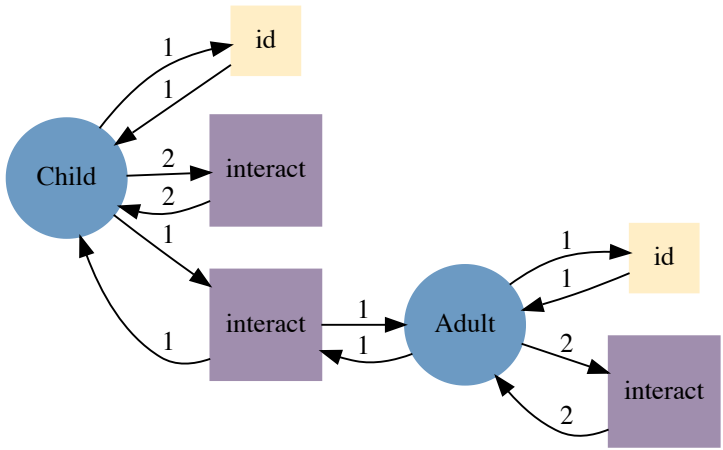

Our final two stratification schemes are spatial stratifications with population dynamics, which were studied in Citron et. al, 2021. In the flux model of spatial dynamics individuals move between patches and only individuals in the same patch can interact.

In [17]:
```
flux_metapopulation = LabelledPetriNet([:Visit1, :Visit2],
    :travel => (:Visit1 => :Visit2),
    :travel => (:Visit2 => :Visit1),
    :id => (:Visit1 => :Visit1),
    :id => (:Visit2 => :Visit2),
    :interact => ((:Visit1, :Visit1) => (:Visit1, :Visit1)),
    :interact => ((:Visit2, :Visit2) => (:Visit2, :Visit2))
)

typed_flux = ACSetTransformation(flux_metapopulation, infectious_type,
  S = [s, s],
  T = [t_strata, t_strata, t_disease, t_disease, t_interact, t_interact],
  I = [i_strata, i_strata, i_disease, i_disease, i_interact1, i_interact2, i_interact1, i_interact2],
  O = [o_strata, o_strata, o_disease, o_disease, o_interact1, o_interact2, o_interact1, o_interact2],
  Name = name -> nothing # specify the mapping for the loose ACSet transform
);

Graph_typed(typed_flux)
```

Out[17]:

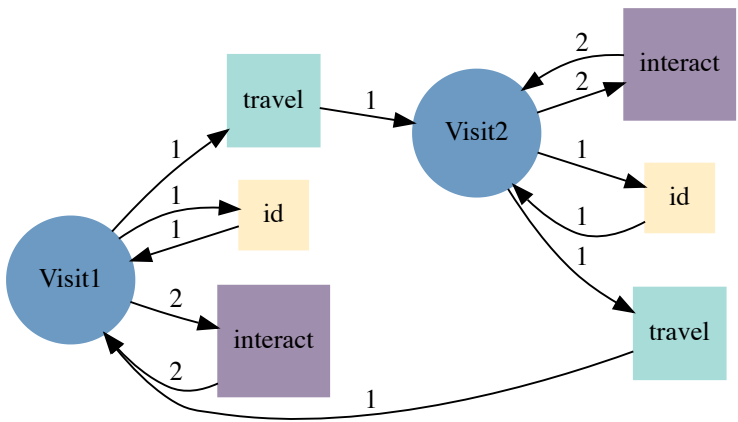

In the simple trip model of spatial dynamics, individuals are associated to two patches: a "residence" patch and a "currently visiting" patch. Individuals can spontaneously move between visiting a patch and returning home, but they cannot change which patch they live in. Only individuals visiting the same patch can interact. This model involves $N^2$ places and $O(N^2)$ transitions where $N$ is the number of patches. To reduce the complexity of building this model, we build it as the model of where individuals live *stratified* by the flux model already implemented above. This example highlights that stratified models can be built *hierarchically*, meaning that a stratification scheme can itself be a stratified model.

33

In [18]:
```
living_model = LabelledPetriNet([:Live1, :Live2],
    :id_disease => (:Live1, :Live1),
    :id_disease => (:Live2, :Live2),
    :id_strata => (:Live1, :Live1),
    :id_strata => (:Live2, :Live2),
    :interact => ((:Live1, :Live1) => (:Live1, :Live1)),
    :interact => ((:Live2, :Live2) => (:Live2, :Live2)),
    :interact => ((:Live1, :Live2) => (:Live1, :Live2)),
    :interact => ((:Live2, :Live1) => (:Live2, :Live1))
)

typed_living = ACSetTransformation(living_model, infectious_type,
  S = [s, s],
  T = [t_disease, t_disease, t_strata, t_strata, t_interact, t_interact, t_interact, t_interact],
  I = [i_disease, i_disease, i_strata, i_strata, i_interact1, i_interact2, i_interact1, i_interact2, i_i
  O = [o_disease, o_disease, o_strata, o_strata, o_interact1, o_interact2, o_interact1, o_interact2, o_i
  Name = name -> nothing # specify the mapping for the loose ACSet transform
);

Graph_typed(typed_living)
```
Out[18]:

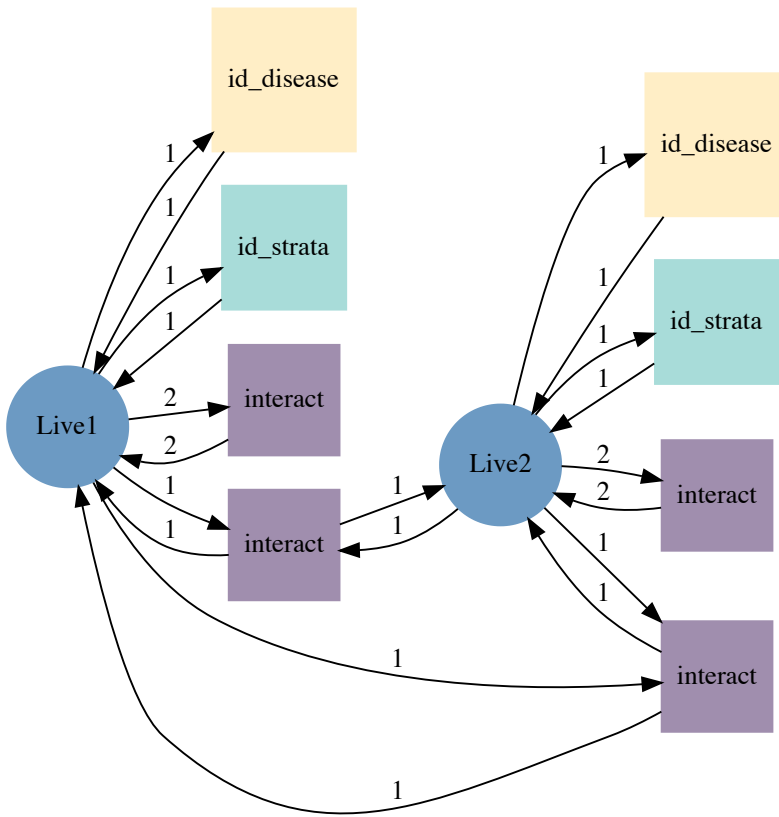

In [19]:
```
typed_simple_trip = typed_stratify(typed_flux, typed_living)

simple_trip = stratify(typed_flux, typed_living)

Graph_typed(typed_simple_trip)
```
Out[19]:

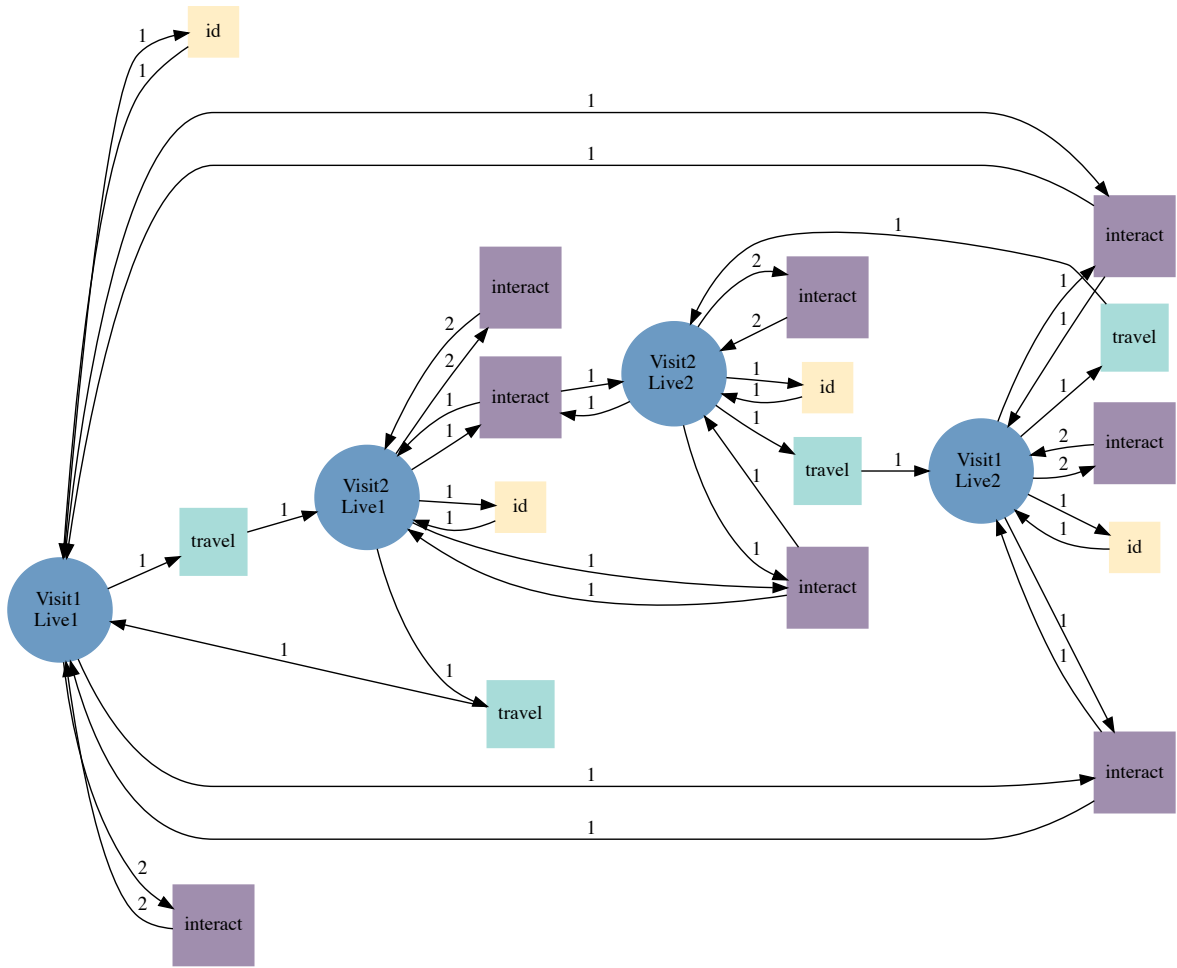

Finally, we define our palette of stratification schemes.

```
In [20]:    stratification_scheme = Dict(
              :quarantine => typed_quarantine,
              :age => typed_age,
              :flux => typed_flux,
              :simple_trip => typed_simple_trip,
            );
```

## Stratified models

Now we can stratify any disease model by any stratification scheme. For example, we can stratify the SVIIvR disease model by the age stratification scheme.

```
In [21]:    typed_model = typed_stratify(disease_models[:SVIIvR], stratification_scheme[:age])
            Graph_typed(typed_model)
```

Out[21]:

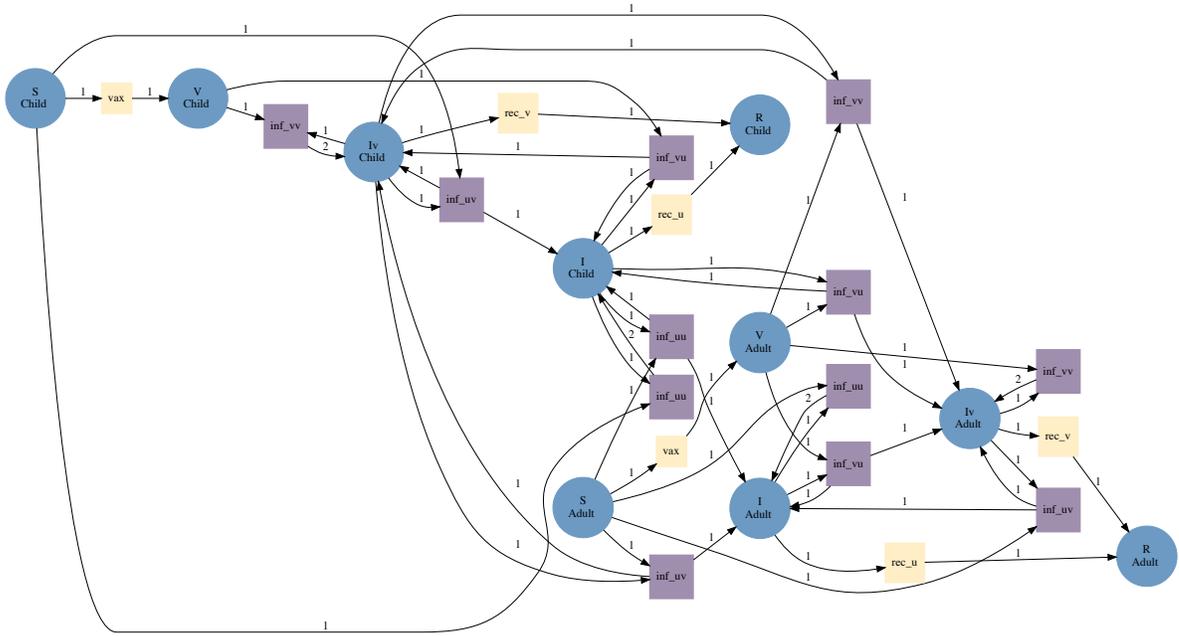

The process of model stratification automates the labor intensive approach taken in Citron et. al, 2021 for combining candidate disease models with candidate movement models in order to test against real-world data.

Below we show how the Petri nets for SIR and SIS disease dynamics stratified by the flux and simple trip movement models are automatically generated by our approach. When the law of mass action is applied to these Petri nets, we recover the ODEs in Equations 6, 7, 9, and 10 of Citron et. al, 2021 modulo the treatment of total vs. proportional populations.

In [22]:
```
typed_SIR_flux = typed_stratify(disease_models[:SIR], stratification_scheme[:flux])
Graph_typed(typed_SIR_flux)
```

Out[22]:
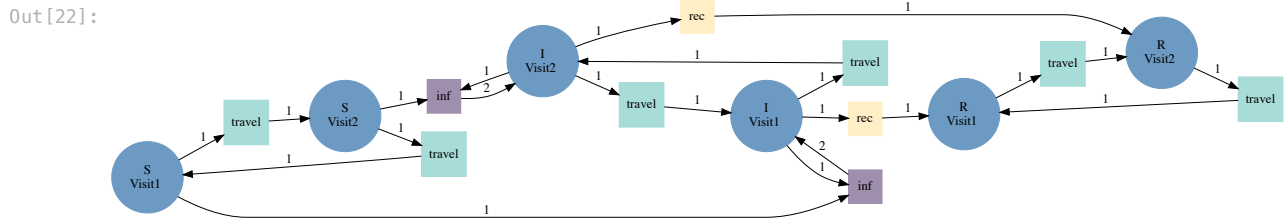

In [23]:
```
typed_SIS_flux = typed_stratify(disease_models[:SIS], stratification_scheme[:flux])
Graph_typed(typed_SIS_flux)
```

Out[23]:
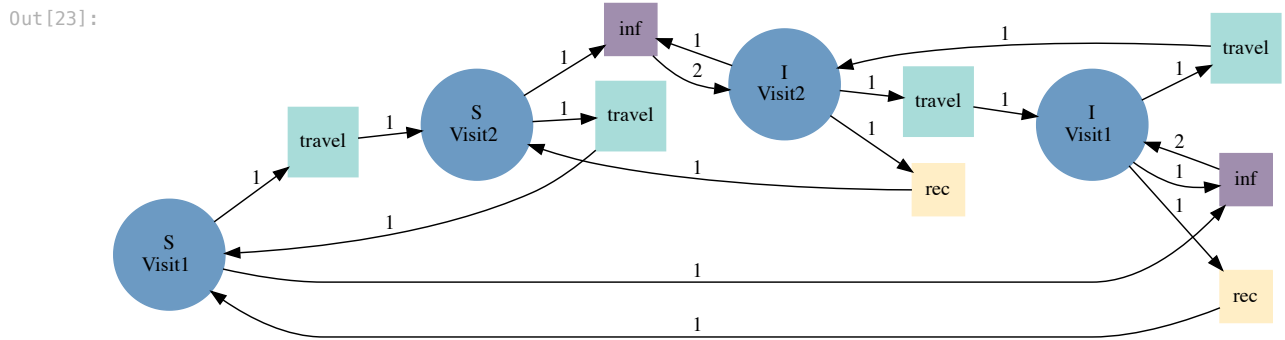

In [24]:
```
typed_SIR_simple_trip = typed_stratify(disease_models[:SIR], stratification_scheme[:simple_trip])
Graph_typed(typed_SIR_simple_trip)
```

Out[24]:

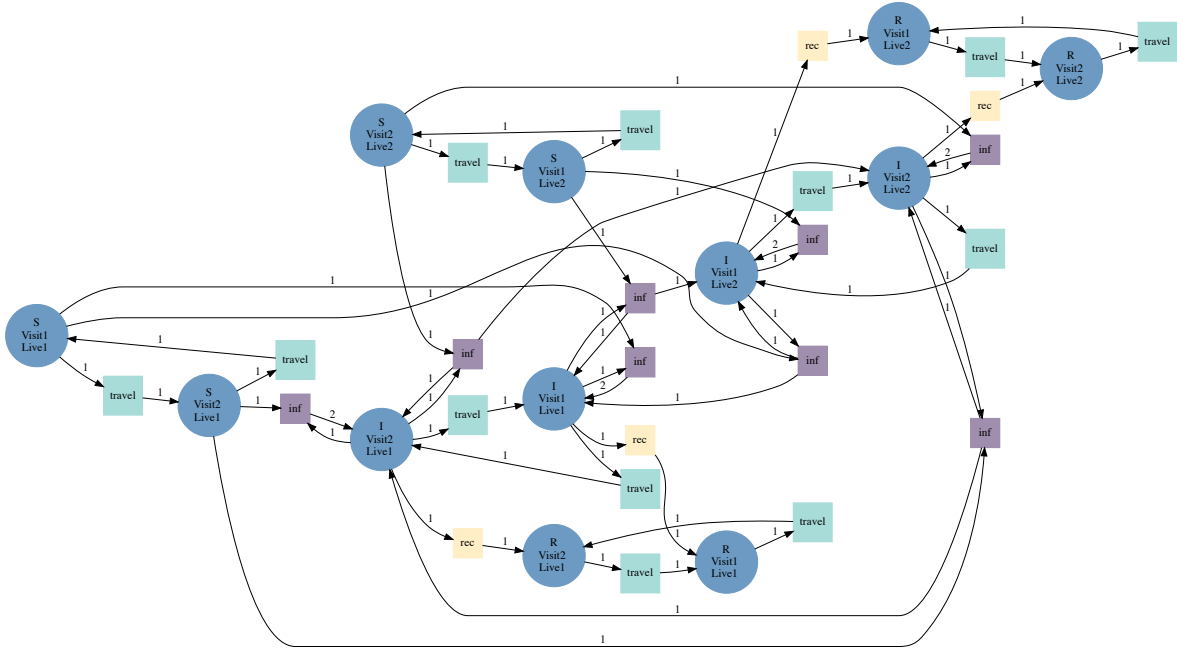

In [25]:
```
typed_SIS_simple_trip = typed_stratify(disease_models[:SIS], stratification_scheme[:simple_trip])
Graph_typed(typed_SIS_simple_trip)
```

Out[25]:

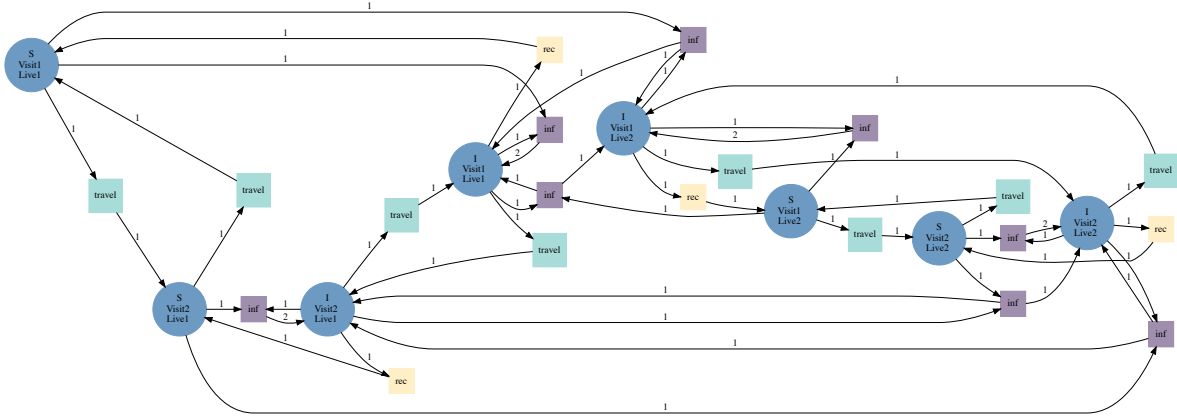

The SIR and SIS simple trip models are called multiply-stratified or hierarchically-stratified models because the simple trip model is itself a stratified model. Fully decomposed, the SIS simple trip model is a 3-way stratification of the SIS disease model, the flux model, and a model of where individuals live. An important property guaranteed by the categorical formalism is that the order of stratification does not affect the final model. For example, the typed Petri net below is isomorphic to the typed Petri net given in `typed_SIS_simple_trip` above.

In [26]:
```
typed_SIS_simple_trip2 = typed_stratify(
  typed_stratify(disease_models[:SIS], stratification_scheme[:flux]),
  typed_living
)
Graph_typed(typed_SIS_simple_trip2)
```

Out[26]:

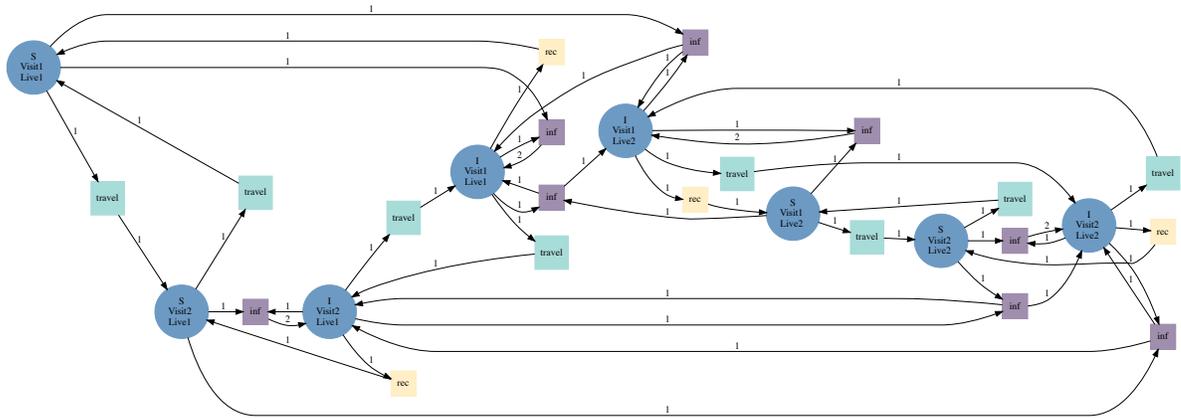

## References

Citron, Daniel T., et al. "Comparing metapopulation dynamics of infectious diseases under different models of human movement." *Proceedings of the National Academy of Sciences* 118.18 (2021).



\end{document}